\documentclass[a4paper,fleqn,usenatbib]{mnras}

\usepackage{newtxtext,newtxmath}

\usepackage[T1]{fontenc}
\usepackage{ae,aecompl}
\usepackage{pdflscape}

\usepackage{graphicx}
\usepackage{grffile}
\usepackage{amsmath}	
\usepackage{amssymb}	

\title[Physical conditions of ionized gas]{Disentangling the physical parameters of gaseous nebulae and galaxies}

\author[D.~Kashino \& A.~K.~Inoue]{
Daichi Kashino,$^{1}$\thanks{E-mail: kashinod@phys.ethz.ch}
Akio K. Inoue,$^{2}$
\\
$^{1}$Department of Physics, ETH Z{\" u}rich, Wolfgang-Pauli-strasse 27, CH-8093 Z{\" u}rich, Switzerland\\
$^{2}$Department of Environmental Science and Technology, Faculty of Design Technology, Osaka Sangyo University, \\3-1-1, Nakagaito, Daito, Osaka 574-8530, Japan\\
}

\date{Accepted XXX. Received YYY; in original form ZZZ}

\pubyear{2015}

\begin{document}
\label{firstpage}
\pagerange{\pageref{firstpage}--\pageref{lastpage}}
\maketitle

\begin{abstract}

We present an analysis to disentangle the connection between physical quantities that characterize the conditions of ionized H\,{\sc ii} regions -- metallicity ($Z$), ionization parameter ($U$), and electron density ($n_\mathrm{e}$) -- and the global stellar mass ($M_\ast$) and specific star formation rate ($\mathrm{sSFR}=\mathrm{SFR}/M_\ast$) of the host galaxies.  We construct composite spectra of galaxies at $0.027 \le z \le 0.25$ from Sloan Digital Sky Survey, separating the sample into bins of $M_\ast$ and sSFR, and estimate the nebular conditions from the emission line flux ratios.  Specially, metallicity is estimated from the direct method based on the faint auroral lines [O\,{\sc iii}]$\lambda$4363 and [O\,{\sc ii}]$\lambda\lambda$7320,7330.  The metallicity estimates cover a wide range from $12+\log\mathrm{O/H}\sim7.6\textrm{--}8.9$.  It is found that these three nebular parameters all are tightly correlated with the location in the $M_\ast$--sSFR plane.  With simple physically-motivated ans{\" a}tze, we derive scaling relations between these physical quantities by performing multi regression analysis.  In particular, we find that $U$ is primarily controlled by sSFR, as $U \propto \mathrm{sSFR}^{0.43}$, but also depends significantly on both $Z$ and $n_\mathrm{e}$.  The derived {\it partial} dependence of $U \propto Z^{-0.36}$ is weaker than the {\it apparent} correlation ($U\propto Z^{-1.52}$).  The remaining negative dependence of $U$ on $n_\mathrm{e}$ is found to be $U \propto n_\mathrm{e}^{-0.29}$.  The scaling relations we derived are in agreement with predictions from theoretical models and observations of each aspect of the link between these quantities.  Our results provide a useful set of equations to predict the nebular conditions and emission-line fluxes of galaxies in semi-analytic models.

\end{abstract}



\section{Introduction}
\label{sec:intro}

Ionized nebulae in star-forming galaxies show a wide range of physical conditions.  The properties of H\,{\sc ii} regions are characterized by a set of physical quantities, including density (or pressure), gas-phase metallicity (hereafter metallicity or $Z$), and the ionization parameter $U$.  These parameters are connected through various fundamental physical processes taking place in the gas, such as photoionization, collisional excitation, and radiative cooling.  Interestingly, the properties of the interstellar medium (ISM) appear also strongly correlated with the global properties of the host galaxies, such as stellar mass ($M_\ast$) and star formation rate (SFR).  This implies that the history and the on-going activities of the galaxies affect the conditions and the processes occurring in the resident nebulae.  Disentangling the connections between these parameters is thus essentially important to understand the astrophysical processes govern the evolution and behavior of baryons in the galaxies.  

The correlation between $M_\ast$ and $Z$ (the ``mass--metallicity'' (MZ) relation) has been well established at low redshifts ($z\sim0.1$) based on a large sample from the Sloan Digital Sky Survey (SDSS) \citep[e.g.,][]{2004ApJ...613..898T}, and also at higher redshifts ($z\gtrsim1$) as well \citep[e.g.,][]{2006ApJ...644..813E,2011ApJ...730..137Z,2012PASJ...64...60Y,2013ApJ...763...92Z,2014ApJ...792...75Z,2015ApJ...799..138S,2017ApJ...835...88K,2018arXiv181201529K}.  \citet{2008ApJ...672L.107E} pointed out the presence of an anticorrelation between $Z$ and SFR at fixed $M_\ast$, which has then further explored by many authors \citep[e.g.,][]{2010A&A...521L..53L,2012MNRAS.422..215Y,2013MNRAS.434..451L,2010MNRAS.408.2115M,2013ApJ...765..140A}.  In particular, \citet{2010MNRAS.408.2115M} proposed the so-called fundamental metallicity relation (FMR) in the $M_\ast$-SFR-$Z$ space.  In these studies, the scatter of the MZ relation is explained by the result of variations in the inflowing rate of circumgalactic medium into the system; inflowing gas drives formation of new stars while diluting the gas-phase metallicity.  In this context, it has also been argued that the relation between $M_\ast$, $Z$ and the gas content is more {\it fundamental}, rather than SFR \citep[e.g.,][]{2013MNRAS.433.1425B,2016MNRAS.455.1156B,2016A&A...595A..48B}.   However, the shape of the FMR and its redshift evolution are still under debate \citep[e.g.,][]{2012MNRAS.422..215Y,2013ApJ...765..140A,2016ApJ...823L..24K,2016ApJ...827...35T}.

A difficulty in measuring metallicity is that in most cases we cannot use the ``direct'' method and thus must rely on the so-called ``strong-line'' method.  The direct method utilizes the flux ratio between auroral and strong lines, such as [O\,{\sc iii}]$\lambda$4363/[O\,{\sc iii}]$\lambda$5007, to measure the electron temperature $T_\mathrm{e}$ of the ionized gas.  Once the temperature is determined, the absolute oxygen abundance can be estimated from the $T_\mathrm{e}$-dependent ratios of strong lines, such as [O\,{\sc ii}]$\lambda$3727 and [O\,{\sc iii}]$\lambda$5007, relative to the H$\beta$ line.  On the other hand, the strong-line method is a technique that uses a relation between strong line ratios (e.g., $R_\mathrm{23}=(\textrm{[O\,{\sc ii}]}\lambda3727+\textrm{[O\,{\sc iii}]}\lambda4959,5007)/\mathrm{H\beta}$) and metallicity, calibrated theoretically or observationally \citep[e.g.,][]{2002ApJS..142...35K,2004MNRAS.348L..59P,2006A&A...459...85N,2008A&A...488..463M,2008ApJ...681.1183K,2013A&A...559A.114M}.  Since the detection of the auroral lines is usually challenging due to their faintness, especially at high metallicity, the strong-line method has been routinely used.  However, the strong-line ratios are functions not only of metallicity, but also of other parameters such as the ionization parameter.  Such systematic uncertainties could result in a misleading comparison between different galaxy populations, e.g., low and high redshifts galaxies \citep[e.g.,][]{2015ApJ...812L..20K,2016ApJ...827...35T}.

The electron density $n_\mathrm{e}$ of the ionized gas is routinely estimated from the intensity ratio of the [S\,{\sc ii}]$\lambda$6717,6731 or [O\,{\sc ii}]$\lambda$3726,3729 doublet lines.  The typical electron densities of H\,{\sc ii} regions in low redshift galaxies have been measured to be $n_\mathrm{e}\sim50\textrm{--}120~\mathrm{cm^{-3}}$ \citep{2008MNRAS.385..769B}.  The densities in high redshift ($z\gtrsim1\textrm{--}2$) galaxies, however, have found to often be elevated relative to low redshifts, ranging from $10^2$ to a few $\times 10^3~\mathrm{cm^{-3}}$ \citep[e.g.,][]{2014ApJ...787..120S,2014ApJ...785..153M,2015MNRAS.451.1284S,2016ApJ...816...23S,2017ApJ...835...88K,2017MNRAS.465.3220K}.  In particular, \citet{2015MNRAS.451.1284S} found that $n_\mathrm{e}$ is correlated with sSFR and/or SFR surface density.  Such correlation could naturally explain the observed evolution in the typical $n_\mathrm{e}$ with redshift.

Higher ionization parameters are often measured in high redshift and/or high sSFR galaxies \citep[e.g.,][]{2009ApJ...701...52H,2014MNRAS.442..900N,2015ApJ...801...88S,2017ApJ...836..164S}.  These studies routinely use the [O\,{\sc iii}]$\lambda$5007/[O\,{\sc ii}]$\lambda$3727 ratio as a proxy of $U$.  Its effectiveness has been established theoretically \citep[e.g.,][]{2000ApJ...542..224D,2002ApJS..142...35K}, while it is known that the line ratio is also sensitive to metallicity at fixed $U$.  Pioneering studies \citep{1985ApJS...58..125E,1986ApJ...307..431D} claimed the presence of a strong inverse correlation between $U$ and $Z$ from comparisons between their theoretical photoionization model and observations. \citet{2006ApJ...647..244D} have theoretically shown that stellar ionizing photon flux illuminating the gas reduces at higher metallicity because of increasing absorption due to a higher opacity in the stellar winds, and due to more efficient conversion from radiation energy into kinetic energy of the winds.  Recent observations of high redshift galaxies (hence higher sSFR and lower $Z$) have also found anticorrelations between $U$ and $Z$ via the strong-line method \citep{2014MNRAS.442..900N,2016ApJ...822...42O}.  In contrast, however, \citet{2018MNRAS.tmp..962K} recently concluded that the ionization parameter $U$ is directly linked to sSFR, while little or probably not with metallicity, from comparison between samples of local and $z\sim1.5$ star-forming galaxies.  It remains under debate whether $U$ and $Z$ are intrinsically correlated or not, and if it is the case, what physical mechanisms work behind the relationship.

The physical quantities that characterize the conditions of galaxies and the ionized nebulae are connected with each other in complicated ways.  As mentioned above, each aspect of this big puzzle has been studied by many authors.  Now we need to disentangle their connection from a viewpoint beyond focusing on a correlation between two or three quantities simultaneously.  In this paper, we will focus on the integrated global properties of galaxies (i.e., $M_\ast$ and SFR) and the basic parameters that define the conditions of the ionized gas locally; metallicity $Z$ (O/H), ionization parameter $U$, and electron density $n_\mathrm{e}$.  To achieve the goal, accurate and precise measurements of oxygen abundance is essentially important, and thus we use the direct method that requires a detection of faint auroral lines of oxygen ions such as [O\,{\sc iii}]$\lambda$4363, and [O\,{\sc ii}]$\lambda$7320, 7330.  To overcome the faintness of these lines, we employ techniques presented in \citet{2013ApJ...765..140A} and \citet{2017MNRAS.465.1384C}, who stacked thousands of galaxy spectra from SDSS.  This enables us to use the direct method even for low SFR and metal-rich galaxies.

This paper is organized as follows.  We describe our sample of galaxies and the methods of spectral stacking and emission-line measurement in Section \ref{sec:data}.  Derivation of the physical properties of ionized nebulae (i.e., $Z$, $U$ and $n_\mathrm{e}$) is described in Section \ref{sec:params}.  In Section \ref{sec:analyses}, we describe the set of simple, physically-motivated ans{\" a}tze for the relations between these quantites, and analyses of the correlations of the stacked measurements.  We present results in Section \ref{sec:results}, and discussions in Section \ref{sec:discussions}.  Summary and the conclusions are given in Section \ref{sec:conclusions}.  Throughout the paper, the stellar masses and SFRs are computed by using a \citet{2001MNRAS.322..231K} universal initial mass function (IMF).  The metallicity $Z$ denotes the gas-phase oxygen abundance, which is defined as the abundance ratio of oxygen atoms to $10^{12}$ hydrogen atoms, as $12+\log (\mathrm{O/H})$.  We adopt the solar metallicity to be $12+\log (\mathrm{O/H})_\odot=8.69$, or $Z_\odot=0.014$ for metallic mass fraction \citep{2009ARA&A..47..481A}.

\section{Data and method}
\label{sec:data}

\subsection{Sample selection}
\label{sec:sample}

The parent sample of galaxies used for this study was constructed following the method described in \citet{2013ApJ...765..140A}.  We here present a brief overview of the sample selection.  We utilize the public MPA-JHU catalog, which contains $\sim930,000$ galaxies from the SDSS Data Release 7 \citep{2009ApJS..182..543A}.  The MPA-JHU catalog contains for each galaxy the estimates of stellar mass and total SFR from the aperture-corrected H$\alpha$ luminosity, with a Kroupa IMF \citep{2004MNRAS.351.1151B,2007ApJS..173..267S}.

\begin{figure*}
	\includegraphics[width=7in]{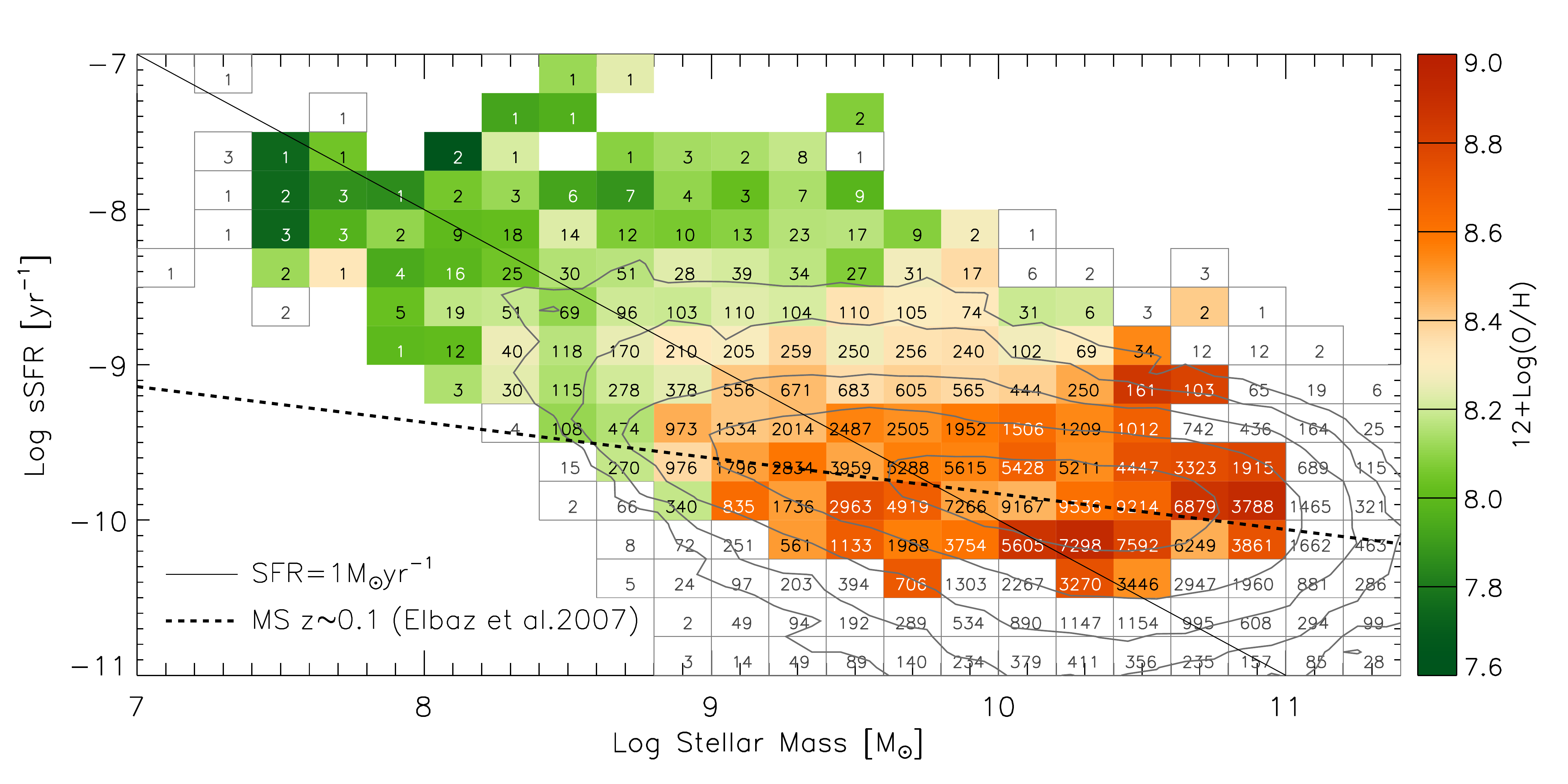}
	\caption{Stellar mass $M_\ast$ vs. sSFR, color-coded with the metallicity determined via the direct method for each stack represented by boxes.  The number of galaxies is reported in each bin.  The dotted line indicates the power-law regression of the star-forming main sequence derived by \citet{2007A&A...468...33E}.  The solid line corresponds to a constant SFR of $1~M_\odot~\mathrm{yr^{-1}}$.  Non-colored boxes indicate stacks for which neither [O\,{\sc iii}] nor [O\,{\sc ii}] auroral lines are detected, and thus are not used for analyses in this paper (except an analysis of the Balmer absorption presented in Appendix \ref{sec:BalmerAbsorption}).}
    \label{fig:M_vs_SFR_metal}
\end{figure*}

The galaxies are limited to have a spectroscopic redshift between $0.027 \le z \le 0.25$ with an uncertainty of $\sigma_z<0.001$.  For this redshift range, both the most blue [O\,{\sc ii}]$\lambda3727$ line and the most red [O\,{\sc ii}]$\lambda\lambda$7320,7330 lines fall within the observed wavelength range of 3800--9200~\AA.  We further imposed on the sample the detection of H$\beta$, H$\alpha$, and [N\,{\sc ii}]$\lambda$6584 all at $S/N \ge 5$.  Possible AGNs were removed by using the Baldwin-Phillips-Terlevich (BPT; \citealt{1981PASP...93....5B}; see also \citealt{1987ApJS...63..295V}) diagram with an empirical classification curve derived by \citet{2003MNRAS.341...33K}.  Objects with possible poor constraints of photometry were removed, based on the warning flags ({\tt DEBLEND\_NOPEAK} or {\tt DEBLENDED\_AT\_EDGE}), or visual inspection.  The final sample consists of 194,005 galaxies and the median redshift is $z=0.081$.  Figure \ref{fig:M_vs_SFR_metal} show the distribution of our selected galaxies in the $M_\ast$--sSFR plane by contours.  The ridge line of the contours is well matched to the so-called main sequence of star-forming galaxies derived by \citet{2007A&A...468...33E}.

\subsection{Spectral stacking}

To determine the metallicity via the direct method, it is required to measure the fluxes of the faint auroral emission lines of oxygen ions,  [O\,{\sc iii}]$\lambda$4363 and [O\,{\sc ii}]$\lambda\lambda$7320, 7330. However, the intensity of these lines rapidly declines below the noise level of the single spectrum as the metallicity increases.  To probe a wide range of the parameter space, we improved the S/N ratios by creating composite spectra of galaxies.  

Our motivation is to disentangle the connections between physical parameters characterizing local conditions in ionized nebulae and global properties of host galaxies.  The stellar mass and SFR are both {\it extensive} variables, and hence are naturally correlated with each other.  The interpretation would be more clarified by using $M_\ast$ as a single indicator of the size of the system, while taking sSFR, an {\it intensive} variable, rather than SFR.  We therefore separated the sample galaxies into the grid in the $M_\ast$--sSFR plane with binsizes of 0.20~dex in $M_\ast$ and 0.25~dex in sSFR.  Figure \ref{fig:M_vs_SFR_metal} shows the numbers of galaxies in each box that represents a single stack.

We utilized the public spectral data from the SDSS DR7, which have been processed with the SDSS Spec2D pipeline \citep{2002AJ....123..485S}.  Before creating composite spectra, the individual spectra were 1) corrected for the Milky Way reddening with the extinction values from \citet{1998ApJ...500..525S} and adopting a \citet{1989ApJ...345..245C} extinction law, 2) shifted to the rest frame based on the spectroscopic redshift, 3) resampled with a universal wavelength grid with a spacing of $\Delta \log_{10} \lambda = 5 \times 10^{-5}$, and 4) normalized by the H$\beta$ flux.  The individual spectra have then been co-added by taking the mean flux between the 25th and the 75th percentiles at each wavelength grid, similarly to \citet{2017MNRAS.465.1384C}.

\subsection{Stellar continuum subtraction}
\label{sec:stesub}

To measure the fluxes of the nebular emission lines, it is important to fit and subtract the stellar component from the composite spectra.  In particular, the Balmer lines and [O\,{\sc iii}]$\lambda$4363 in the vicinity of H$\gamma$, are highly impacted by the stellar atmospheric absorption.  For the stellar continuum subtraction, we generated a synthetic spectrum for each composite spectrum.  We performed the fit using the IDL version of the pPXF package \citep{2017MNRAS.466..798C} with the built-in library of single stellar population (SSP) spectral templates.  The library includes 150 model spectra based on the MILES library of stellar templates \citep{2006MNRAS.371..703S}, computed for 25 different ages and 6 different metallicities using the code presented in \citet{2010MNRAS.404.1639V}.  The templates cover a wavelength range of 3540--7410~\AA\ with a resolution of 2.51\AA\ (full width at half maximum; FWHM).

\begin{figure*}
	\includegraphics[width=6in]{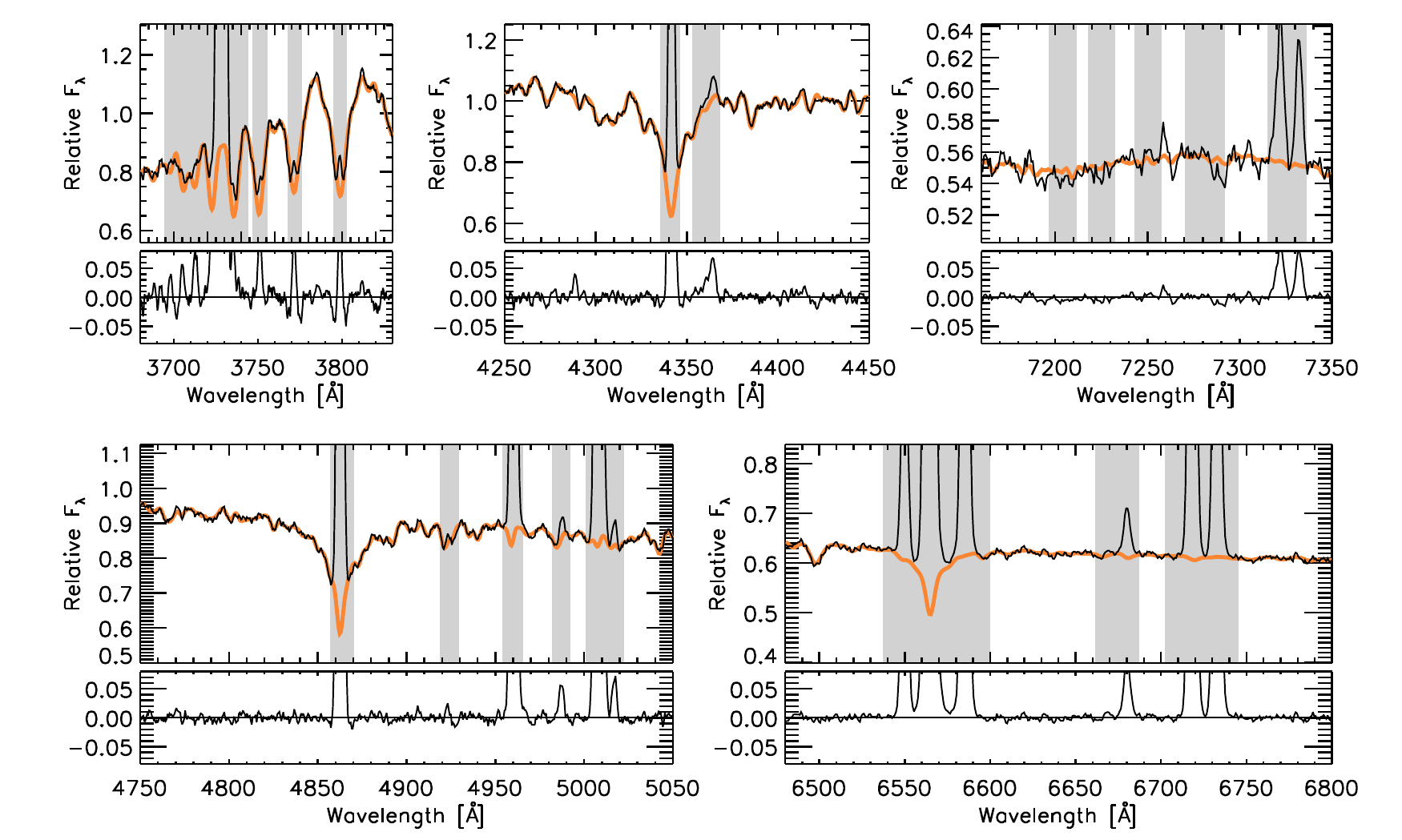} 
	\caption{Example composite spectrum for a stack of $\log M_\ast = \left[ 9.0:9.2\right]$, $\log \mathrm{sSFR}=\left[-9.25:-9.00\right]$ within the wavelength ranges around the key emission lines; [O\,{\sc ii}]$\lambda\lambda$3727,3729 (upper left-hand panel), H$\gamma$ and [O\,{\sc iii}]$\lambda$4363 (upper middle panel), H$\beta$ and [O\,{\sc iii}]$\lambda\lambda$4959,5007 (lower left-hand panel), H$\alpha$, [N\,{\sc ii}]$\lambda\lambda 6548,6584$, and [S\,{\sc ii}]$\lambda\lambda6717,6731$ (lower right-hand panel), and [O\,{\sc ii}]$\lambda\lambda$7320,7330 (upper right-hand panel).  For each window, the composite spectra (black solid line) and the best-fit stellar template (orange thick line) are shown in the main part and the residuals after stellar subtraction is shown in the lower part of each panel.  Gray stripes indicate the ranges that were masked out for the stellar template fitting.  Note that, in the [O\,{\sc ii}]$\lambda\lambda$7320,7330 window, we masked out some wavelength ranges where features are found in some spectra (but not unambiguously identified as being attributed to specific species)}.
    \label{fig:specimg}
\end{figure*}

We performed the stellar template fitting first using the entire spectral range between 3650~\AA\ and 7360~\AA\ to determine the best-fit synthesized spectrum, generated as an arbitrary superposition of the 150 simple stellar population spectra in the library.  We then fit the single best-fit template in each subrange of a few hundred angstroms around the emission lines of interest, allowing slight modifications on the systemic offset, velocity dispersion, and additive and multiplicable factors to improve the quality of the fits.  In each spectral subrange, narrow wavelength regions where the emission lines present were masked out.  In Figure \ref{fig:specimg}, we show an example of spectral stacking and stellar continuum subtraction.  The panels correspond to the spectral ranges for the key emission lines, [O\,{\sc ii}]$\lambda\lambda$3726,3729, H$\gamma$, [O\,{\sc iii}]$\lambda$4363, H$\beta$, [O\,{\sc iii}]$\lambda\lambda$4959,5007, H$\alpha$, [N\,{\sc ii}]$\lambda\lambda$6548,6584, [S\,{\sc ii}]$\lambda\lambda$6717,6731, and [O\,{\sc ii}]$\lambda\lambda$7320,7330, respectively.  

Figure \ref{fig:specimg} shows that the absorption features are prominent at wavelengths of the Balmer series lines.  These Balmer absorption lines arise from the absorption of photons by hydrogen excited at the $n=2$ level in the photospheres of stars.  This effectively reduces the observed fluxes of the Balmer lines if one simply excludes the continuum based on a linear or low-order polynomial fitting around the lines.  Meanwhile, it is not always possible to perform the stellar continuum subtraction based on the stellar population synthesis fitting, as done in this paper.  Therefore, it would be useful to derive a simple empirical relation between the effect of Balmer absorption and the galaxy properties.   In Appendix \ref{sec:BalmerAbsorption}, we present the amount of the Balmer absorption as a function of $M_\ast$ and sSFR, and provide empirical relations for absorption correction. 

\subsection{Line flux measurement}

We performed emission-line fitting using the MPFIT package for IDL \citep{2009ASPC..411..251M} to measure the fluxes of the emission lines in the stellar-subtracted composite spectra.   Each emission line was fit with a single gaussian profile.  Note that all the doublet lines ([O\,{\sc ii}]$\lambda\lambda$3726,3729, [O\,{\sc iii}]$\lambda\lambda$4959,5007, [N\,{\sc ii}]$\lambda\lambda6548,6584$, [S\,{\sc ii}]$\lambda\lambda6717,6731$, and [O\,{\sc ii}]$\lambda\lambda7320,7330$) were fit with independent amplitudes.  
We disregarded lines that were poorly fit with low S/N ($<5$), which is computed for each line as a ratio of measured flux and the uncertainty based on the errors on the fitting parameters returned by the MPFIT procedure.  The measured fluxes of all emission lines were then corrected for dust extinction by adopting a \citet{1989ApJ...345..245C} extinction law. The absolute level of extinction was estimated based on the observed H$\alpha$/H$\beta$ ratio by assuming the intrinsic values of H$\alpha$/H$\beta$ given in \citet[][Table 4-4]{2006agna.book.....O} for Case B recombination with $n_\mathrm{e}=10^2~\mathrm{cm^{-3}}$.  The intrinsic H$\alpha$/H$\beta$ ratio is very insensitive to the electron density if $n_\mathrm{e}<10^4~\mathrm{cm^{-3}}$, while being slightly dependent of the electron temperature.  For the range of $7000 \lesssim T_\mathrm{e}(\mathrm{O^+}) \lesssim 20000~\mathrm{K}$ of our data, the H$\alpha$/H$\beta$ ratio could vary within $\pm 5\%$, increasing as $T_\mathrm{e}$ decreases.  Therefore we iteratively estimated the extinction as follows.  We first estimate the extinction by assuming the intrinsic H$\alpha$/H$\beta$ ratio of 2.86 for $T_\mathrm{e}=10^4~\mathrm{K}$ and derive the electron temperature from the extinction-corrected fluxes.  The estimated $T_\mathrm{e}$ is then used to modify the intrinsic H$\alpha$/H$\beta$ ratio, and to re-compute the level of extinction.  We then repeat the extinction correction of all the line fluxes and derivation of the physical quantities.  The change in the resulting $\log \mathrm{O/H}$ after the iteration is correlated with $\log \mathrm{O/H}$, and is negligible at $\log \mathrm{O/H}\lesssim8.2$ while gradually being larger at higher $Z$.  The final estimates after the iteration are reduced by $\approx0.08~\mathrm{dex}$ at $12+\log \mathrm{O/H}=8.8$ relative to the initial estimates with the intrinsic H$\alpha$/H$\beta$ of 2.86 assumed.

As pointed out by \citet{2013ApJ...765..140A}, an emission feature was found at $\approx$4360~\AA, which is blended with the [O\,{\sc iii}]$\lambda$4363 auroral line.  \citet{2017MNRAS.465.1384C} suggested that this feature is reasonably associated with emission lines from Fe\,{\sc ii} ions.  Following their treatment, we simultaneously fit the three line components of the possible [Fe\,{\sc ii}] lines that are centered on 4357.7, 4358.8, and 4360.6~\AA, in addition to H$\gamma$ and [O\,{\sc iii}]$\lambda$4363. As shown by \citet{2017MNRAS.465.1384C}, we found that the contribution from the possible [Fe\,{\sc ii}] lines increases as metallicity increases.  Following \citet{2017MNRAS.465.1384C}, we disregarded possible detection of [O\,{\sc iii}]$\lambda$4363 when the flux measured for the $\lambda$4360 feature is $>0.5$ times the flux measured for [O\,{\sc iii}]$\lambda$4363.

\section{Physical parameters}
\label{sec:params}

In this work, we investigate the connection among the five physical quantities: two galaxy global properties, $M_\ast$ and sSFR, and three nebular parameters characterizing the conditions of the H\,{\sc ii} regions, metallicity $Z$ ($12+\log (\mathrm{O/H})$), ionization parameter $U$, and electron density $n_\mathrm{e}$.  These nebular properties were derived based on the stacked measurements of the emission-line fluxes as described in the following subsections.  For this purpose, we used an package of IDL routines called impro\footnote{The code has been developed by John Moustakas et al. and is available here: https://github.com/moustakas/impro}.  This software offers functions similar to the commonly used IRAF/TEMDEN package.  To evaluate the nebular parameters and their uncertainties we generated 200 random realizations for each set of the line fluxes measured from each stack while incorporating the flux errors with Gaussian distributions.  We then took the median and the standard deviation for each parameter estimate from the resultant realizations.  

\subsection{Electron temperature and density}

The electron temperatures $T_\mathrm{e} (\mathrm{O}^{+})$ and $T_\mathrm{e}(\mathrm{O}^{++})$ were determined from the line ratios [O\,{\sc ii}]$\lambda\lambda$3726,3729/[O\,{\sc ii}]$\lambda\lambda$7320,7330 and [O\,{\sc iii}]$\lambda\lambda$4959,5007/[O\,{\sc iii}]$\lambda$4363, respectively.  If we detected both auroral lines, we independently estimated $T_\mathrm{e} (\mathrm{O}^{+})$ and $T_\mathrm{e}(\mathrm{O}^{++})$ using the impro routines.  However, we failed to detect [O\,{\sc iii}]$\lambda$4363 for stacks with an estimated metallicity of $12+\log \mathrm{O/H} \gtrsim 8.3$ due to the flux being faint and the contamination of the $\lambda 4360$ feature being significant.  For these stacks, we estimated the temperature $T_\mathrm{e}(\mathrm{O}^{++})$ by utilizing the so-called flux-flux (ff) relation derived by \citet{2006MNRAS.367.1139P}, following the method in \citet{2017MNRAS.465.1384C}.  This empirically-calibrated relation predicts the [O\,{\sc iii}]$\lambda$4363 flux relative to H$\beta$ from the strong-line ratios, [O\,{\sc iii}]$\lambda$4959,5007/H$\beta$ and [O\,{\sc ii}]$\lambda$3727/H$\beta$ (see Equation (14) of \citealt{2006MNRAS.367.1139P}).  Similarly, for stacks where we failed to detect the [O\,{\sc ii}]$\lambda\lambda$7320,7330 lines, we utilized the empirical relation using the strong-line ratios given by \citet[][see Equation 3]{2017MNRAS.465.1384C}.  In this work, we use the stacks where at least we successfully detect either the [O\,{\sc iii}] or [O\,{\sc ii}] auroral lines, and disregarded stacks having neither detections of these.  We emphasize that, at the metallicity range where the [O\,{\sc iii}]$\lambda$4363 line is undetected, the total oxygen abundance is dominated by the singly-ionized oxygen $\mathrm{O}^{+}$.  This is also the case for the opposite situation, i.e., the oxygen abundance of the stacks without the [O\,{\sc ii}]$\lambda$7320,7330 detection is dominated by $\mathrm{O}^{++}$.  Therefore, the effects of the possible uncertainties from the use of these empirical relations are negligible in the total oxygen abundance estimates.

The flux ratio of the auroral and strong lines depends not only on the electron temperature, but also weakly on the electron density.  The electron density can be estimated from the [S\,{\sc ii}]$\lambda$6717/[S\,{\sc ii}]$\lambda$6731 ratio, with a small dependence on the temperature $T_\mathrm{e}(\mathrm{S}^{+})$.  We therefore iteratively estimated both $T_\mathrm{e}$ and $n_\mathrm{e}$ assuming $T_\mathrm{e}(\mathrm{S}^{+})=T_\mathrm{e}(\mathrm{O}^{+})$.  The [S\,{\sc ii}]  doublet ratio saturates at $n_\mathrm{e}<10~\mathrm{cm^{-3}}$.  As a result, the errors is getting larger for lower $n_\mathrm{e}$, or cannot be constrained for some cases.  In our analysis, we set the lower limit of the $n_\mathrm{e}$ range to be $n_\mathrm{e}=5~\mathrm{cm^{-3}}$.  We substituted $\log n_\mathrm{e}/(\mathrm{cm^{-3}})=0.7\pm0.7$ for the value of three stacks for which the observed [S\,{\sc ii}] ratio is above the saturation limit.  The derived $n_\mathrm{e}$ of the stacks ranges from $\sim17$ to $\sim300~\mathrm{cm^{-3}}$ (the central 90 percentiles), with the median $n_\mathrm{e}=76~\mathrm{cm^{-3}}$.  Note that this range and the median value do not represent the electron densities of the individual SDSS galaxies since the number of galaxies in the stacks are hugely different from one to another bin.  Lastly, we note that the spectral resolution of the SDSS spectra are not sufficient to obtain reasonable estimates of $n_\mathrm{e}$ from the [O\,{\sc ii}]$\lambda\lambda$3726,3729 doublet ratio.

\subsection{Direct abundance determination}
\label{sec:abundance}

The ionic oxygen abundance O$^+$/H$^+$ and O$^{++}$/H$^+$ were determined from [O\,{\sc ii}]$\lambda\lambda$3726,3727/H$\beta$, and [O\,{\sc iii}]$\lambda$5007/H$\beta$ as well as from the temperature estimates, $T_\mathrm{e} (\mathrm{O}^{+})$ and $T_\mathrm{e}(\mathrm{O}^{++})$, respectively.  The procedure in the impro package includes a slight dependence on $n_\mathrm{e}$ as well.  As described above, if we have constraint only either $T_\mathrm{e} (\mathrm{O}^{+})$ or $T_\mathrm{e}(\mathrm{O}^{++})$, we estimated the other using the empirical relations based on the strong lines.  We remind that, for such cases, the contribution of the ions whose auroral line(s) is not detected is subdominant.  We then assumed that the total oxygen abundance is the sum of the ionic abundances of these two ions, 
\begin{equation}
\frac{\mathrm{O}}{\mathrm{H}}=\frac{\mathrm{O}^+}{\mathrm{H^+}}+\frac{\mathrm{O}^{++}}{\mathrm{H}^+}.
\end{equation}
We neglected the minimal contribution of neutral atoms since our attention is focused on the conditions of the H\,{\sc ii} regions, where the neutral fraction is typically small ($\sim10^{-4}$).  The contribution of $\mathrm{O^{+++}}$ and higher ionization ions was also neglected, whose ionization potential is 54.9~eV or higher, thus there is little chance to be ionized by stellar radiation.

\subsection{Ionization parameter}
\label{sec:ionizationparameter}

The ionization parameter is a key quantity in this work.  It is locally defined as the ratio of ionizing photon flux $S_\mathrm{ion}$ and the hydrogen density $n_\mathrm{H}$: $q_\mathrm{ion} = S_\mathrm{ion}/n_\mathrm{H}$.  The ionization parameter can be made dimensionless by dividing by the speed of light as follows:
\begin{equation}
U = q_\mathrm{ion}/c = n_\gamma / n_\mathrm{H}
\label{eq:U}
\end{equation}
where $n_\gamma$ is the volume density of the ionizing photons.  Throughout the paper, we use the dimensionless $U$, instead of $q_\mathrm{ion}$.  We note that the electron density $n_\mathrm{e}$ is almost equivalent to $n_\mathrm{H}$ in a usual H\,{\sc ii} region.

We estimated $U$ from a sensitive indicator, the flux ratio of [O\,{\sc iii}]$\lambda$5007/[O\,{\sc ii}]$\lambda$3726,3729 \citep{2000ApJ...542..224D,2002ApJS..142...35K}.  Although the flux ratio is also sensitive to metallicity, we can resolve this degeneracy using the metallicity estimate from the direct method (Section \ref{sec:abundance}).  The [O\,{\sc iii}]/[O\,{\sc ii}] ratio also slightly varies with $n_\mathrm{e}$ at fixed $U$ and $Z$, as well.  We consider this effect by using the $n_\mathrm{e}$ estimate from the [S\,{\sc ii}] doublet ratio.

We determined the ionization parameters for each stack by interpolating the flux ratios computed for grids of $Z$, $U$, and $n_\mathrm{H}$, assuming $n_\mathrm{H}=n_\mathrm{e}$.  We used the CLOUDY (C17.00) photoionization code to compute the emission-line fluxes for given nebular conditions.  For the gas phase, we adopted the solar chemical composition, though the chemical composition does little affect the estimates of $U$ since it is estimated from the flux ratio of the same species at different ionization levels (i.e., O$^+$ and O$^{++}$).  Photoionization calculations were executed in the pseudo plane-parallel geometry (i.e., inner radius $\gg$ thickness of the ionized shell) for the following values of $Z$, $U$, and $n_\mathrm{H}$.
\begin{eqnarray}
12+\log (\mathrm{O/H}) &=& 7.49, 7.69, 7.89, 8.09, 8.29, 8.39, \nonumber \\
                                      && 8.49, 8.59, 8.69, 8.79 8.99, 9.19\nonumber \\
\log U &=& -4.00, -3.75, -3.50, -3.25, -3.00 \nonumber \\
&& -2.75, -2.50, -2.25, -2.00 \nonumber \\
\log n_\mathrm{H} /(\mathrm{cm^{-3}}) &=& 0.5, 1.0, 1.5, 2.0, 2.5, 3.0 \nonumber
\end{eqnarray}
Ionizing spectra used in our CLOUDY calculations were produced by the Starburst99 population synthesis code \citep{1999ApJS..123....3L}, adopting an constant SFR history with a \citet{2001MNRAS.322..231K} IMF ($0.1\textrm{--}100~M_\odot$) and the Geneva high mass-loss evolutionary tracks with five values of stellar metallicity: $Z_\ast=0.001$, 0.004, 0.008, 0.020, and 0.040, corresponding to $12+\log(\mathrm{O/H})=7.54\textrm{--}9.15$.  We adopted a set of spectra at $10~\mathrm{Myr}$, and, for a given gas-phase $Z$, the input ionizing spectrum was produced by interpolating the spectral set at $Z_\ast = Z$ to match the stellar and gas-phase metallicities.  The ionization parameter $U$ is then derived from a $U$--[O\,{\sc iii}]/[O\,{\sc ii}] relation obtained by interpolating the three-dimensional grid data at the estimated $n_\mathrm{H}=n_\mathrm{e}$ and $\log\mathrm{O/H}$.  The estimates of $U$ range from $\log U = -3.5$ to $-2.4$.
  
\subsection{Derived quantities}
\label{sec:derivation}

\begin{figure}
	\includegraphics[width=3.5in]{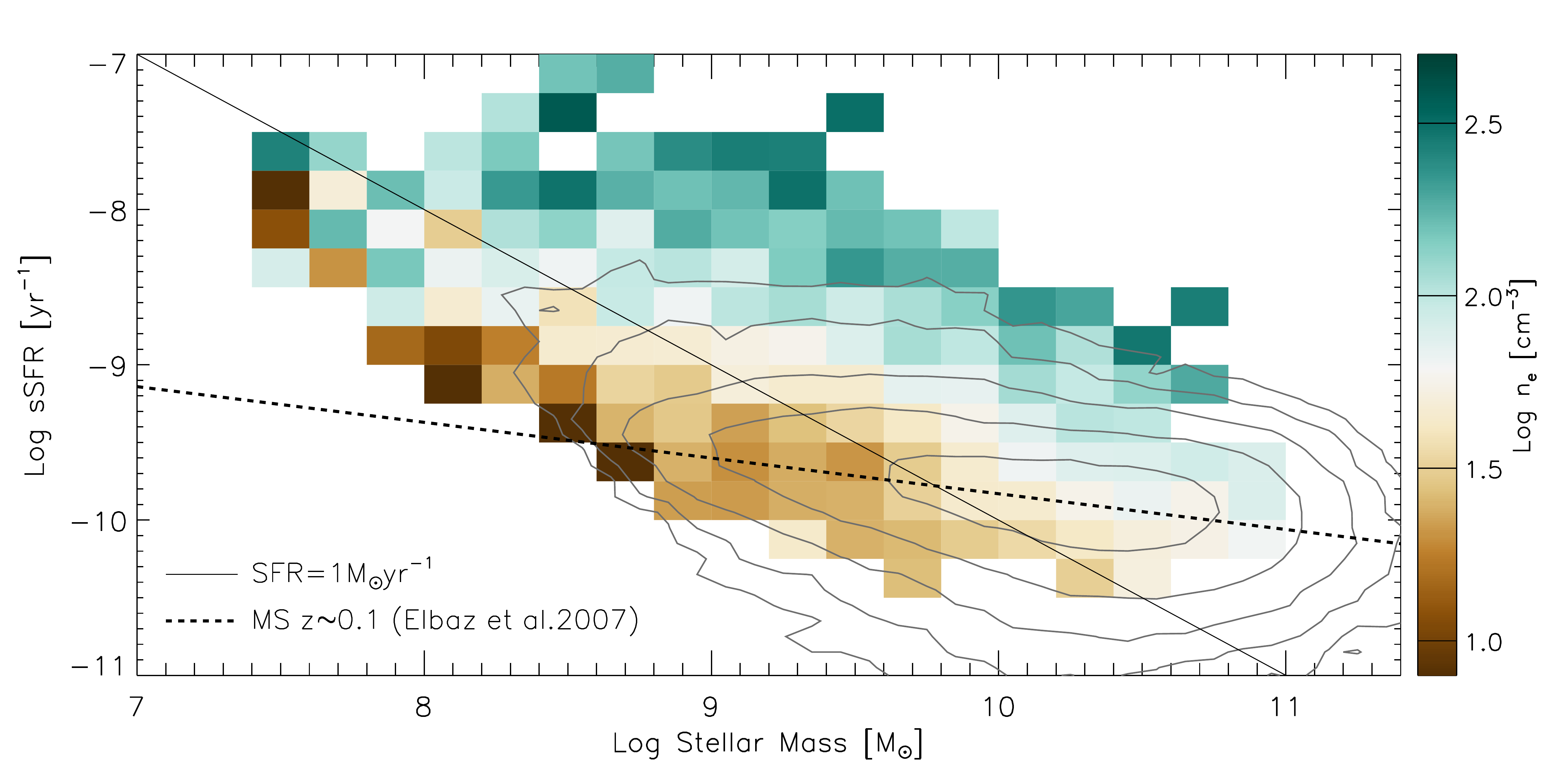}
	\includegraphics[width=3.5in]{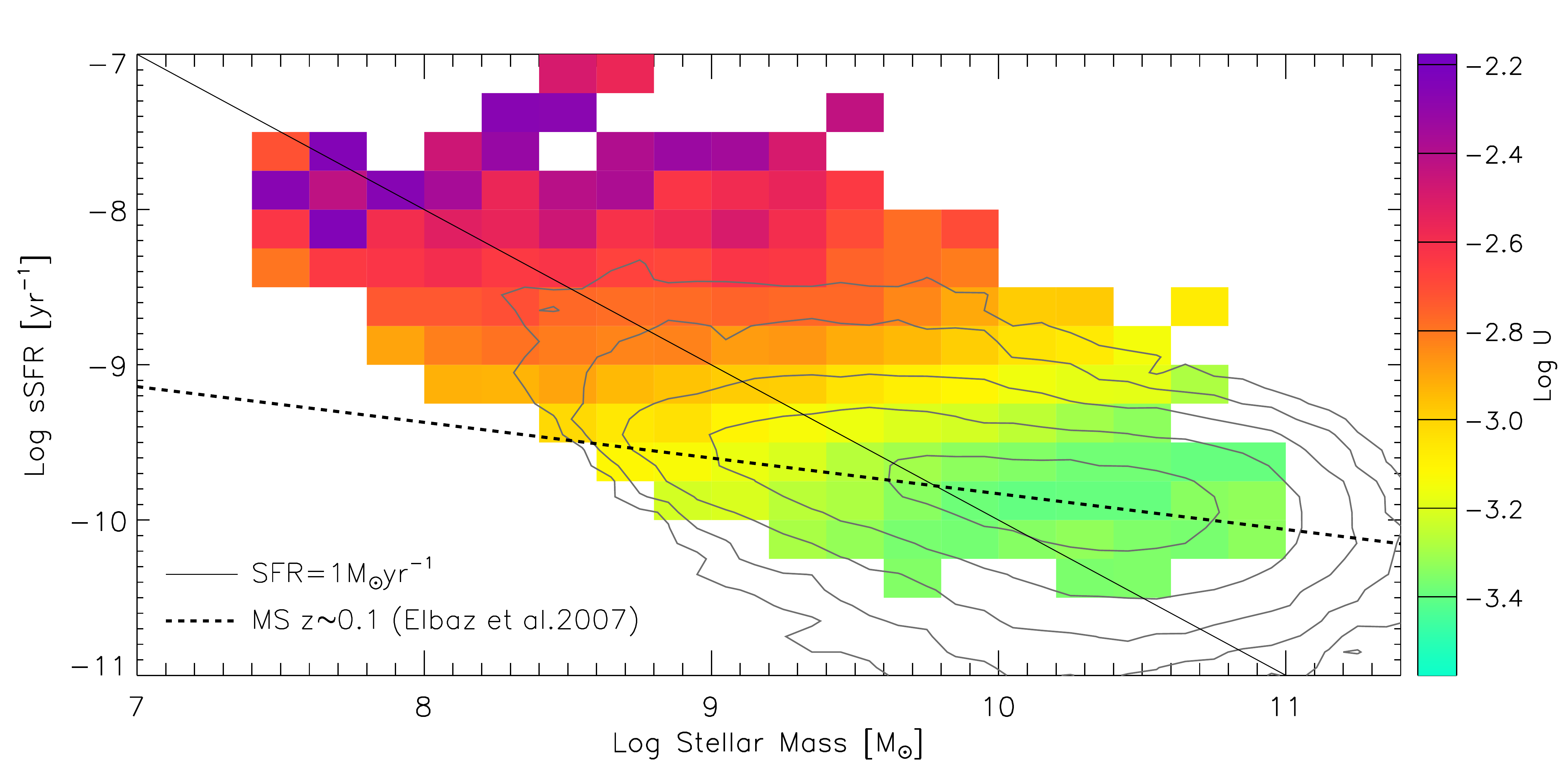}
	\caption{Stellar mass $M_\ast$ vs. sSFR, color-coded by the electron density $n_\mathrm{e}$ (upper panel) and by the ionization parameter $U$ (lower panel) for each stack. The contours and lines are the same as in Figure \ref{fig:M_vs_SFR_metal}.}
    \label{fig:M_vs_SFR_others}
\end{figure}

We successfully estimated the metallicity for 138 stacks, of which we detected both [O\,{\sc iii}]$\lambda$4363 and [O\,{\sc ii}]$\lambda\lambda$7320,7330 for 75 stacks, only [O\,{\sc ii}]$\lambda\lambda$7320,7330 for 61 stacks, and only [O\,{\sc iii}]$\lambda$4363 for 2 stacks.  Figure \ref{fig:M_vs_SFR_metal} shows the derived metallicity by colors in each stacking bin in the $M_\ast$--sSFR plane.  Here a caveat is that the stacks with successful metallicity estimation are biased towards lower $M_\ast$ and higher SFR, resulting in being biassed to the population above the main sequence (dotted line; \citealt{2007A&A...468...33E}).  This is because the intensity of auroral lines rapidly declines with increasing $Z$ as well as with decreasing SFR.  As a result, the detection of the auroral lines has often failed at high $M_\ast (\gtrsim10^{10.6}~M_\odot)$, or below the main sequence even if thousands of spectra are stacked.  In contrast, the detection succeeded at lower $M_\ast$ and high SFRs even with a few (or only a single) galaxies in the stacks.  However, we would emphasize that the successful stacks cover the whole range of the metallicity of interest ($7.6 \lesssim 12+\log (\mathrm{O/H}) \lesssim 8.9$), probed by most studies of the MZ relation at low and higher redshifts (references in Section \ref{sec:intro}), including both regimes of extremely metal-poor galaxy populations \citep[e.g.,][]{2016ApJ...819..110S} and the super-solar.

\begin{table*}
	\caption{Derived physical quantities of the stacks}
	\label{tb:data}
	\begin{tabular}{cccccccccc} 
 \hline
\multicolumn{2}{c}{$M_\ast$ bin} & \multicolumn{2}{c}{sSFR bin} & $N_\mathrm{gal}$ & $\log M_\ast (M_\odot)$ & $\log \mathrm{sSFR} (\mathrm{yr^{-1}})$ & $12+\log(\mathrm{O/H})$ & $\log U$ & $\log n_\mathrm{e} (\mathrm{cm^{-3}})$ \\
(1) & (2) & (3) & (4) & (5) & (6) & (7) & (8) & (9) & (10) \\
 \hline
$ 7.4$ & $ 7.6$ & $ -8.50$ & $ -8.25$ & $   2$ & $ 7.556\pm 0.160$ & $ -8.348\pm  0.169$ & $8.121\pm0.057$ & $ -2.795\pm  0.014$ & $1.913\pm0.164$ \\
$ 7.4$ & $ 7.6$ & $ -8.25$ & $ -8.00$ & $   3$ & $ 7.533\pm 0.127$ & $ -8.115\pm  0.132$ & $7.735\pm0.063$ & $ -2.631\pm  0.015$ & $1.073\pm0.810$ \\
$ 7.4$ & $ 7.6$ & $ -8.00$ & $ -7.75$ & $   2$ & $ 7.512\pm 0.053$ & $ -7.983\pm  0.077$ & $7.748\pm0.017$ & $ -2.264\pm  0.010$ & $0.870\pm0.706$ \\
$ 7.4$ & $ 7.6$ & $ -7.75$ & $ -7.50$ & $   1$ & $ 7.587\pm 0.082$ & $ -7.641\pm  0.136$ & $7.744\pm0.091$ & $ -2.716\pm  0.025$ & $2.427\pm0.371$ \\
$ 7.6$ & $ 7.8$ & $ -8.50$ & $ -8.25$ & $   1$ & $ 7.798\pm 0.075$ & $ -8.476\pm  0.120$ & $8.329\pm0.073$ & $ -2.642\pm  0.025$ & $1.306\pm0.693$ \\
$ 7.6$ & $ 7.8$ & $ -8.25$ & $ -8.00$ & $   3$ & $ 7.693\pm 0.067$ & $ -8.101\pm  0.082$ & $7.966\pm0.031$ & $ -2.245\pm  0.017$ & $2.229\pm0.677$ \\
$ 7.6$ & $ 7.8$ & $ -8.00$ & $ -7.75$ & $   3$ & $ 7.650\pm 0.051$ & $ -7.839\pm  0.064$ & $7.861\pm0.022$ & $ -2.426\pm  0.010$ & $1.694\pm0.662$ \\
$ 7.6$ & $ 7.8$ & $ -7.75$ & $ -7.50$ & $   1$ & $ 7.601\pm 0.092$ & $ -7.542\pm  0.151$ & $8.043\pm0.029$ & $ -2.250\pm  0.014$ & $2.108\pm0.377$ \\
$ 7.8$ & $ 8.0$ & $ -9.00$ & $ -8.75$ & $   1$ & $ 7.994\pm 0.133$ & $ -8.754\pm  0.268$ & $7.996\pm0.063$ & $ -2.894\pm  0.021$ & $1.166\pm0.761$ \\
$ 7.8$ & $ 8.0$ & $ -8.75$ & $ -8.50$ & $   5$ & $ 7.941\pm 0.032$ & $ -8.574\pm  0.060$ & $8.027\pm0.033$ & $ -2.732\pm  0.011$ & $1.943\pm0.175$ \\
$ 7.8$ & $ 8.0$ & $ -8.50$ & $ -8.25$ & $   4$ & $ 7.973\pm 0.033$ & $ -8.366\pm  0.071$ & $7.942\pm0.083$ & $ -2.629\pm  0.029$ & $2.180\pm0.523$ \\
$ 7.8$ & $ 8.0$ & $ -8.25$ & $ -8.00$ & $   2$ & $ 7.897\pm 0.082$ & $ -8.200\pm  0.094$ & $8.103\pm0.027$ & $ -2.593\pm  0.009$ & $1.799\pm0.453$ \\
$ 7.8$ & $ 8.0$ & $ -8.00$ & $ -7.75$ & $   1$ & $ 7.864\pm 0.140$ & $ -7.958\pm  0.155$ & $7.852\pm0.022$ & $ -2.261\pm  0.013$ & $2.214\pm0.397$ \\
$ 8.0$ & $ 8.2$ & $ -9.25$ & $ -9.00$ & $   3$ & $ 8.181\pm 0.046$ & $ -9.109\pm  0.159$ & $8.151\pm0.118$ & $ -2.925\pm  0.026$ & $0.718\pm0.714$ \\
$ 8.0$ & $ 8.2$ & $ -9.00$ & $ -8.75$ & $  12$ & $ 8.171\pm 0.026$ & $ -8.822\pm  0.070$ & $8.007\pm0.058$ & $ -2.819\pm  0.019$ & $1.040\pm0.704$ \\
$ 8.0$ & $ 8.2$ & $ -8.75$ & $ -8.50$ & $  19$ & $ 8.125\pm 0.030$ & $ -8.586\pm  0.053$ & $8.177\pm0.031$ & $ -2.733\pm  0.008$ & $1.660\pm0.276$ \\
$ 8.0$ & $ 8.2$ & $ -8.50$ & $ -8.25$ & $  16$ & $ 8.130\pm 0.024$ & $ -8.331\pm  0.039$ & $7.982\pm0.021$ & $ -2.594\pm  0.008$ & $1.825\pm0.289$ \\
$ 8.0$ & $ 8.2$ & $ -8.25$ & $ -8.00$ & $   9$ & $ 8.135\pm 0.029$ & $ -8.134\pm  0.041$ & $8.001\pm0.017$ & $ -2.525\pm  0.008$ & $1.487\pm0.683$ \\
$ 8.0$ & $ 8.2$ & $ -8.00$ & $ -7.75$ & $   2$ & $ 8.164\pm 0.089$ & $ -7.856\pm  0.099$ & $8.056\pm0.021$ & $ -2.360\pm  0.012$ & $1.967\pm0.382$ \\
$ 8.0$ & $ 8.2$ & $ -7.75$ & $ -7.50$ & $   2$ & $ 8.049\pm 0.129$ & $ -7.556\pm  0.133$ & $7.617\pm0.035$ & $ -2.463\pm  0.011$ & $2.003\pm0.589$ \\
$ 8.2$ & $ 8.4$ & $ -9.25$ & $ -9.00$ & $  30$ & $ 8.352\pm 0.021$ & $ -9.073\pm  0.063$ & $8.238\pm0.064$ & $ -2.931\pm  0.012$ & $1.378\pm0.419$ \\
$ 8.2$ & $ 8.4$ & $ -9.00$ & $ -8.75$ & $  40$ & $ 8.309\pm 0.015$ & $ -8.885\pm  0.042$ & $8.259\pm0.031$ & $ -2.791\pm  0.007$ & $1.250\pm0.472$ \\
  \hline
 \multicolumn{10}{p{14cm}}{{\bf Notes.}(1--2) lower and upper $\log M_\ast$ limits of the stack.  (3--4) lower and upper $\log\mathrm{sSFR}$ limits of the stack.  (5) number of galaxies in the stack.  (6) average stellar mass of the stack.  (7) average sSFR of the stack. (8--10) estimated metallicity, ionization parameter, and electron density of the stack.  The table is published in its entirety in the machine-readable format.}
	\end{tabular}
\end{table*}

In Figure \ref{fig:M_vs_SFR_others}, the ionization parameter $U$ and electron density $n_\mathrm{e}$ are presented in the same manner as metallicity.  All these panels in Figures \ref{fig:M_vs_SFR_metal} and \ref{fig:M_vs_SFR_others} clearly show that the nebular conditions depend strongly on $M_\ast$ and sSFR.  It is shown that $Z$ and $U$ are more sensitive to sSFR than to $M_\ast$.  Meanwhile, the electron density $n_\mathrm{e}$ increases both with $M_\ast$ at fixed sSFR as well as with sSFR at fixed $M_\ast$.  Table \ref{tb:data} summarize the derived physical properties (full version including emission line flux measurements available online).  The relations between the observed strong-line fluxes and derived physical quantities are presented in Appendix \ref{sec:diagnostic}.

\section{Correlation analysis}
\label{sec:analyses}

\begin{figure*}
	\includegraphics[width=6.5in]{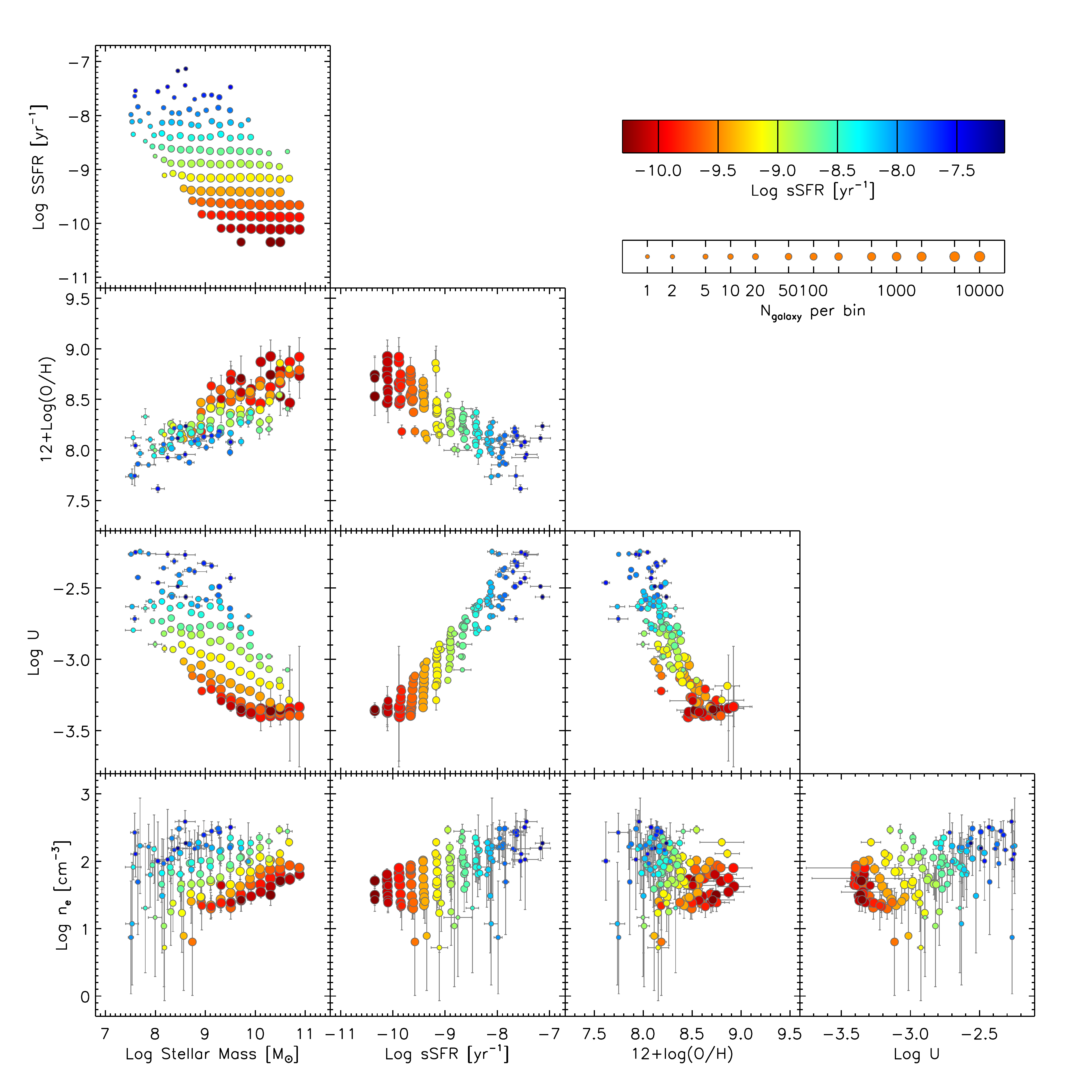}
	\caption{Correlations between two parameters among the five of interest, $\log M_\ast/M_\odot$, $\log \mathrm{sSFR}/(\mathrm{yr^{-1}})$, $12+\log (\mathrm{O/H})$, $ \log U$, and $\log n_\mathrm{e}/(\mathrm{cm^{-3}})$.  Filled circles indicate stacked measurements.  The color and size of the data points indicate the median sSFR and the number of galaxies in each stack, respectively, as shown outside the frame.}
    \label{fig:triangle}
\end{figure*}

In this section, we attempt to disentangle the connection between the parameters of the ionized H\,{\sc ii} regions and the host galaxies.  Figure \ref{fig:triangle} shows correlations for all possible combinations among the five parameters of interest, $\log M_\ast/M_\odot$, $\log \mathrm{sSFR}/(\mathrm{yr^{-1}})$, $12+\log (\mathrm{O/H})$, $ \log U$, and $\log n_\mathrm{e}/(\mathrm{cm^{-3}})$.  The stellar mass and sSFR of each stack are defined as the median values of individual galaxies within the bins.  The data points are color-coded by sSFR, and the symbol size denotes the number of galaxies in each stacking bin, as indicated outside the frames.  

\subsection{Ans{\" a}tze}
\label{sec:ansatze}

\begin{figure}
	\includegraphics[width=3.2in]{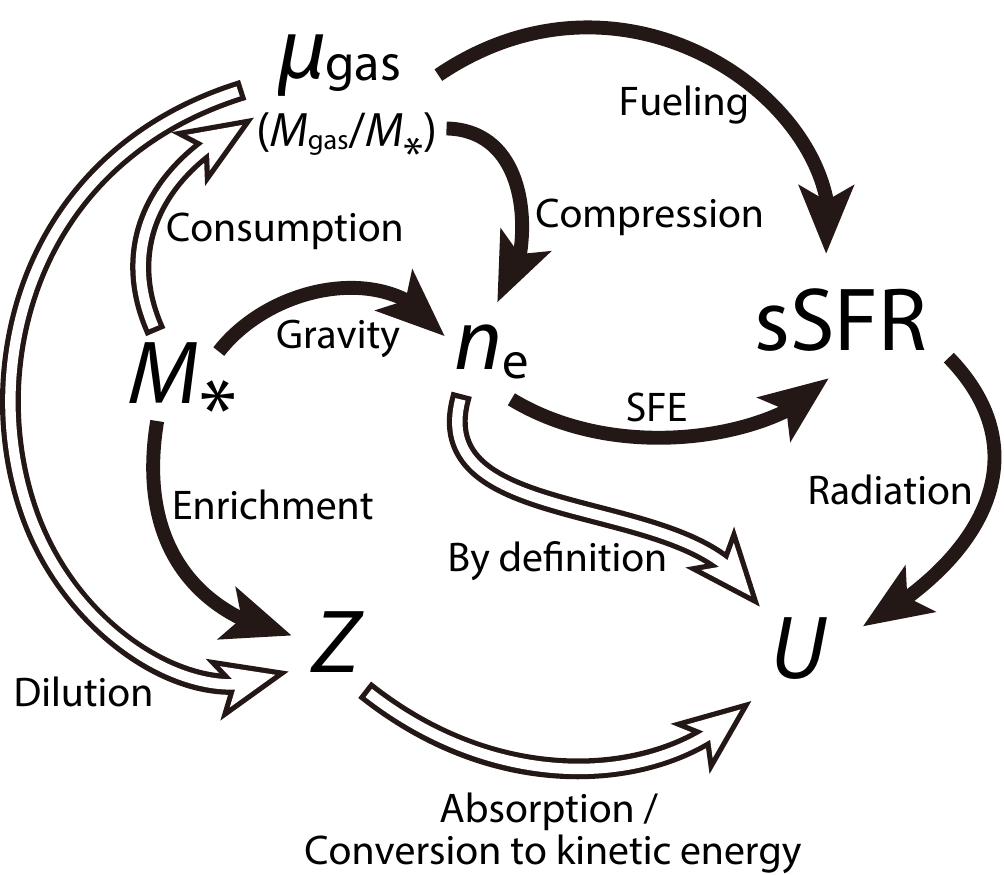}
	\caption{Schematic view of the ans{\" a}z we made in Section \ref{sec:ansatze}.  Key quantities here are the five parameters that we focus on in this paper ($M_\ast$, sSFR, $Z$, $n_\mathrm{e}$, and $U$), as well as an additional parameter, the gas content ($\mu_\mathrm{gas} = M_\mathrm{gas}/M_\ast$).  Black (white) arrows indicate the direction of possible causation which is expected to result in a positive (negative) correlation between the two parameters at their starting and ending points.  Each arrow is labelled with possible explanation of the causality (see text).}
    \label{fig:schem}
\end{figure}

We here propose a set of ans{\" a}tze to figure out the connections between the parameters.  First we assume that the partial correlations between any two of the parameters are expressed by power-law functions, i.e., we assume linearity for their logarithms.  Although this is undoubtedly an oversimplification, it is reasonably supported by the observed relationships shown in Figure \ref{fig:triangle}, and indeed a more elaborate parametrization is not justified by our limited data.  Still, the power-law slopes that we obtain from the data will be helpful to qualitatively understand the physics controlling the conditions of the nebulae and galaxy evolution.

To shape a course to analyze the observed correlations, we employ further assumptions for the dependencies of the parameters.  In Figure \ref{fig:schem}, we schematically show our assumption about the dependencies of the parameters.  The black (white) arrows indicate the direction of possible causation which is expected to lead to a positive (negative) correlation between the two parameters at their starting and ending points.  We here consider for convenience an additional parameter, the gas mass ratio $\mu_\mathrm{gas}=M_\mathrm{gas}/M_\ast$, which is assumed to be uniquely determined once the key parameters of interest are given.

One may naturally argue that the SFR of a galaxy increase with the amount of the fuel, i.e., the gas content in the system.  Star formation consumes the gas reservoir while accumulating the stellar mass of the galaxy.  Thus, the gas mass ratio is naturally expected to be anticorrelated with $M_\ast$.  These causations are illustrated, with labels ``Fueling'' and ``Consumption'' respectively, in Figure \ref{fig:schem}.

Mathematically, the SFR integrated over a galaxy is related to its total molecular gas mass as $\mathrm{SFR} = M_\mathrm{gas}/\tau_\mathrm{dep}$ where $\tau_\mathrm{dep}$ is the gas depletion timescale, and the star formation efficiency is defined as its inverse $\mathrm{SFE} = 1/\tau_\mathrm{dep} = \mathrm{SFR}/M_\mathrm{gas}$.  On the other hand, a simple hypothesis relates the local SFR volume density $\dot{\rho}_\ast$ with the gas volume density $\rho_\mathrm{gas}$ and free-fall timescale $\tau_\mathrm{ff}$ of an overdense region as $\dot{\rho}_\ast \propto \rho_\mathrm{gas}/\tau_\mathrm{ff} \propto \rho_\mathrm{gas}^{1.5}$ since $\tau_\mathrm{ff}\propto \rho_\mathrm{gas}^{1/2}$ \citep[e.g.,][]{1959ApJ...129..243S,1998ApJ...498..541K}.  Although bridging local and global scales is not straightforward, one may naively argue that the global scale depletion time $\tau_\mathrm{dep}$ (or SFE) scales as the local free-fall time (or its inverse).   We could therefore suppose that the sSFR is also controlled by density $n_\mathrm{e}$ through varying ``SFE'', as labeled in Figure \ref{fig:schem}.  We note that the electron density $n_\mathrm{e}$ of H\,{\sc ii} regions is here assumed to be proportional to the density $\rho_\mathrm{gas}$ of star-forming molecular gas clouds, though this is not trivial.

Since volume densities are difficult to observationally measure, the relation between the SFR and gas mass densities is often expressed using the corresponding surface densities as $\dot{\Sigma}_\ast  \approx \Sigma_\mathrm{gas}^N$, which is known as the Schmidt-Kennicutt (SK) relation.  Observations have shown that the SFR and gas mass surface densities averaged over a galaxy scale obey the SK relation with $N\approx 1.4$ \citep{1998ApJ...498..541K}.  This indicates that the local gas density of collapsing gas clouds scales with total amount of gas content in the systems.  We would therefore expect an intrinsic positive correlation between $\mu_\mathrm{gas}$ and $n_\mathrm{e}$, which is expected to be positive.  Although the physical mechanism is not fully clear, this possible causation is labeled as ``Compression'' for convenience.  In addition, the gas density may also be affected by stellar mass $M_\ast$ via gravitational effects (labeled as ``Gravity'').  

Turning to metallicity, the well-known MZ relation could be naturally expected as the consequence of the history of mass assembly and accompanying chemical enrichment.  The current metallicity, however, also encloses the history of gas inflows and outflows.  In particular, the inverse correlation seen between $Z$ and SFR at fixed $M_\ast$ has been commonly interpreted as the result of the so-called ``dilution'' effect: enhanced inflow of metal-poor gas in the intergalactic medium would dilute the metallicity in the ISM, while boosting star formation.  This naturally predicts the anticorrelation between $Z$ and the gas content $\mu_\mathrm{gas}$.  In Figure \ref{fig:schem}, these dependencies of $Z$ on stellar mass and the gas content are indicated, respectively, with labels ``Enrichment'' and ``Dilution''.  We note that, due to the intrinsic anticorrelation between $M_\ast$ and $\mu_\mathrm{gas}$ through the gas consuming mass assembly, $Z$ is naturally expected to be anticorrelated with $\mu_\mathrm{gas}$ {\it even if we do not consider the effects of dilution due to an enhanced inflow rate}.  The dilution scenario implicitly assumes that galaxies of the same $M_\ast$ experienced the equivalent history of star formation and created the same amount of metals, and that the dispersions in the current gas content and thus metallicity are caused by recent fluctuations in the inflow rate in a short timescale.  Keeping this in mind, we consider the effect of $\mu_\mathrm{gas}$ on $Z$, especially at fixed $M_\ast$, and thus named the causation just "Dilution" for simplicity.

Given $n_\mathrm{e}$, $Z$, and sSFR, we can then approach the ionization parameter $U$.  Returning to the simple definition of $U \equiv n_\gamma/n_\mathrm{H}$, we can separate the ionization parameter into two controlling factors, the ionizing photon flux illuminating the gas and the local density of the ionized gas.  It would thus be natural to suppose that $U$ is anticorrelated with $n_\mathrm{e}$ when fixing the ionizing radiation (labeled as ``By definition'' in Figure \ref{fig:schem}).   
Concerning the numerator, it may be natural to assume that the ionizing photon flux is primarily controlled by the sSFR (labeled as ``Radiation'' in Figure \ref{fig:schem}).  However, the increase in the galaxy-wide sSFR would not necessarily lead an increase in the ionizing photon flux in a {\it single} H\,{\sc ii} region, while certainly resulting in an increase in the total amount of the ionizing photons in a whole galaxy.  Lastly, we expect that $U$ is also anticorrelated with $Z$.  This is because an increase in metals contained in the stellar atmosphere of the central massive stars would result in a higher optical depth in the stellar winds, more efficiently absorbing ionizing photons \citep[e.g.,][]{2006ApJ...647..244D}.  In addition, at higher metallicity, ionizing photons are more efficiently scattered in the photospheres and then their radiation energy are converted to kinetic energy of the stellar winds.  We may also expect that an increase in the dust amount in the H\,{\sc ii} regions, which is expected to correlate with metallicity, would reduce the ionizing photon flux contributing to ionization \citep{2001ApJ...555..613I,2001AJ....122.1788I}.  The possible causation between $Z$ and $U$ is labeled as ``Absorption/Conversion to kinetic energy'' in Figure \ref{fig:schem}).

With these simple ans{\" a}tze, we attempt to derive the dependencies of sSFR, $Z$, and $U$ on the other parameters.  The ans{\" a}tze denote $\mathrm{sSFR}=\mathrm{sSFR}(\mu_\mathrm{gas}, n_\mathrm{e})$ and $Z=Z(M_\ast, \mu_\mathrm{gas})$.  However, we need to eliminate $\mu_\mathrm{gas}$ because we have no direct information about the gas content ($\mu_\mathrm{gas}$) of the galaxies.  We have assumed that $n_\mathrm{e}$ is uniquely determined by $M_\ast$ and $\mu_\mathrm{gas}$ as shown in Figure \ref{fig:schem}, and thus $\mu_\mathrm{gas}$ is uniquely determined once $M_\ast$ and $n_\mathrm{e}$ are given.  We thus parametrize sSFR and $Z$ as a function of $M_\ast$ and $n_\mathrm{e}$ eliminating $\mu_\mathrm{gas}$, as well as $U$ as a function of sSFR $Z$, and $n_\mathrm{e}$.  In this section, we explicitly argued which positive or negative correlation is expected between two parameters.  In our analysis, however, we did not impose any limits on the power-law indices and their signs in the following analysis.

\subsection{Multi regression analysis}

Following the ans{\" a}tze we made, we derived the relations between the quantities ($M_\ast$, $\mathrm{sSFR}$, $\mathrm{O/H}$, $U$, and $n_\mathrm{e}$).  The physically-motivated relations to be constrained are  expressed as follows:  
\begin{align}
\log \mathrm{sSFR}+9 &= p_0 + p_1 (\log M_\ast - 10) + p_2 (\log n_\mathrm{e}), \label{eq:sSFR(M,ne)0} \\
4+\log \mathrm{O/H} &= q_0 + q_1 (\log M_\ast - 10) + q_2 (\log n_\mathrm{e}), \label{eq:Z(M,ne)0} \\
\log U &= r_0 + r_1 (\log n_\mathrm{e}) \nonumber \\
&\quad + r_2 (4+\log \mathrm{O/H}) + r_3 (\log \mathrm{sSFR}+9). \label{eq:U(Z,ne,sSFR)0}
\end{align}
We note that $M_\ast$, sSFR, and $n_\mathrm{e}$ are given in units of $M_\odot$, $\mathrm{yr^{-1}}$, and $\mathrm{cm^{-3}}$, respectively.  In addition to these, we also derived the empirical functional forms for the three nebular parameters, $Z$, $U$, and $n_\mathrm{e}$, as a linear function of $\log M_\ast$ and $\log \mathrm{sSFR}$.  Our fitting procedure fits a linear surface expressed by these equations to the stacked measurements, accounting for all the measurement errors on the relevant variables.  

Let us consider a general form of these models as $y = p_0 + \sum_{k=1}^{m} p_k x_k$.  We refer to $y$ and $x_k~(k=1,2,...,m)$ as {\it objective} and {\it explanatory} variables, respectively.  We estimate the model parameters (i.e., the normalization and coefficients of Equations (\ref{eq:sSFR(M,ne)0}--\ref{eq:U(Z,ne,sSFR)0}) based on the maximum likelihood estimation.  The likelihood $\mathcal{L}$ is computed for a given model as follows:
\begin{equation}
\mathcal{L} = \prod_{i=1}^{N} \frac{1}{\sqrt{2 \pi \sigma_\mathrm{err,i}^2}} \exp\left[-\frac{(y^\mathrm{obs}_i - y^\mathrm{mod}_i)^2}{2\sigma_{\mathrm{err},i}^2} \right],
\end{equation}
where $N=138$ is the number of stacked measurements.  The error $\sigma_\mathrm{err,i}$ is obtained for each stack by summing in quadrature the uncertainties expected for the objective ($\delta y_i$) and explanatory ($\delta x_{k,i}$) variables as follows:
\begin{equation}
\sigma_{\mathrm{err},i}^2 = \delta y_i^2 + \sum_{k=1}^{m} \left(p_k \delta x_{k,i} \right)^2.
\label{eq:sigma_tot}
\end{equation}
Note that the value $\sigma_{\mathrm{err},i}$ depends on the coefficients of the model.

We sampled the posterior probability distribution of the model parameters by employing a Markov-Chain Monte-Carlo (MCMC) technique, and using the emcee package for Python \citep{2013PASP..125..306F}.  In our fitting procedure, we adopted a uniform prior probability function for each parameter.

\section{Results} 
\label{sec:results}

\begin{table*}
	\centering
	\caption{Coefficients of the linear surface fitting between the parameters$^a$}
	\label{tb:coeff}
	\begin{tabular}{lccccccc} 
		\hline
		& Intercept & $\log M_\ast (M_\odot) -10$ & $\log n_\mathrm{e} (\mathrm{cm^{-3}})$ & $4+\log (\mathrm{O/H})$ & $\log\mathrm{sSFR}(\mathrm{yr^{-1}})+9$ & $\chi^2/\mathrm{dof}$$^b$ & $\sigma_\mathrm{int}$$^c$ \\
		\hline
$\log \mathrm{sSFR}+9$  & $ -3.661^{+  0.124}_{-  0.125}$ & $ -0.627^{+  0.028}_{-  0.029}$ & $  1.753^{+  0.063}_{-  0.064}$ & - & - &    0.99 & - \\
$4+\log (\mathrm{O/H})$ & $  1.238^{+  0.049}_{-  0.046}$ & $  0.228^{+  0.011}_{-  0.011}$ & $ -0.401^{+  0.024}_{-  0.024}$ & - & - &    1.61 & 0.0633 \\
$\log U$                & $ -2.316^{+  0.043}_{-  0.039}$ & - & $ -0.292^{+  0.020}_{-  0.022}$ & $ -0.360^{+  0.037}_{-  0.037}$ & $  0.428^{+  0.014}_{-  0.014}$ &    1.89 & 0.0585 \\
\hline
$\log n_\mathrm{e}$     & $  2.066^{+  0.012}_{-  0.012}$ & $  0.310^{+  0.021}_{-  0.019}$ & - & - & $  0.492^{+  0.020}_{-  0.021}$ &    0.85 & - \\
$4+\log (\mathrm{O/H})$ & $  0.427^{+  0.007}_{-  0.007}$ & $  0.136^{+  0.007}_{-  0.006}$ & - & - & $ -0.208^{+  0.008}_{-  0.008}$ &    5.71 & 0.0998 \\
$\log U$                & $ -3.073^{+  0.002}_{-  0.002}$ & $ -0.137^{+  0.002}_{-  0.002}$ & - & - & $  0.372^{+  0.002}_{-  0.002}$ &   26.07 & 0.0874 \\
\hline
\multicolumn{8}{l}{$^a$ We note that $M_\ast$, sSFR, and $n_\mathrm{e}$ are given in units of $M_\odot$, $\mathrm{yr^{-1}}$, and $\mathrm{cm^{-3}}$, respectively.}\\
\multicolumn{8}{l}{$^b$ Reduced chi-square statistic.}\\
\multicolumn{8}{l}{$^c$ Intrinsic scatter with respect to the relevant objective variable.}\\
	\end{tabular}
\end{table*}

In Table \ref{tb:coeff}, we summarize the coefficients constrained by multi regression analysis.  The upper three rows are the results for the physically-motivated models as defined in Section \ref{sec:ansatze} (Equations \ref{eq:sSFR(M,ne)0}--\ref{eq:U(Z,ne,sSFR)0}).  The lower three rows show the results obtained from regression on $M_\ast$ and sSFR.  The reduced $\chi^2$ statistics are computed with these best-fit parameters as follows:
\begin{equation}
\chi^2/\mathrm{dof} = \sum_\mathrm{i=1}^N \frac{(y^\mathrm{obs}_i - y^\mathrm{model}_i)^2}{\sigma_{\mathrm{err},i}^2}/(N-m-1).
\end{equation}
Note that the $(m+1)$ is the number of parameters to be constrained.  We also computed the residual intrinsic scatter assuming that the individual stacked measurements are intrinsically scattered with respect to the best fit following a Gaussian distribution of $\sigma_\mathrm{int}$.  In practice, we estimated $\sigma_\mathrm{int}$ that satisfies the following relation:
\begin{equation}
\sum_\mathrm{i=1}^N \frac{(y^\mathrm{obs}_i - y^\mathrm{model}_i)^2}{\sigma_{\mathrm{err},i}^2+\sigma_\mathrm{int}^2}/(N-m-1) = 1.
\end{equation}

The reduced $\chi^2$ statistics for the fits with the physically-motivated ans{\"a}tze are reasonably close to unity, with the residual scatter below $0.1$~dex.  This indicates that these models well fit the data with little residual information.  For $Z$ and $U$, the $\chi^2/\mathrm{dof}$ values are larger for the results from the regression on only $M_\ast$ and sSFR.  Generally, the $\chi^2/\mathrm{dof}$ statistics can be used to evaluate which model is better fitting the {\it same} data.  However, in the case here, the $\chi^2/\mathrm{dof}$ values do {\it not} immediately indicate that, for example for $Z$, the fit with $M_\ast$ and $n_\mathrm{e}$ is better than the fit with $M_\ast$ and sSFR because the different models use the different variables.  The intrinsic scatters are anyway smaller than $0.1~\mathrm{dex}$ for the fits with $M_\ast$ and sSFR, similarly to the fits with the ans{\"a}tze.  Hereafter we focus our attention to the results for the physically-motivated models.

We found the best-fit relations for the sSFR, metallicity $Z$, and the electron density $n_\mathrm{e}$, as follows (the errors are given in Table \ref{tb:coeff}).
\begin{alignat}{1}
\log \mathrm{sSFR} &= -12.661 -0.627 \left(\log M_\ast - 10\right) + 1.753 \log n_\mathrm{e}, \label{eq:sSFR(M,ne)} \\
12+\log\mathrm{O/H} &=  9.238 + 0.228 \left( \log M_\ast -10 \right) - 0.401 \log n_\mathrm{e}, \label{eq:Z(M,ne)} \\
\log U &= -2.316 - 0.360 \left( \log \mathrm{O/H} + 4 \right) \nonumber \\
&\quad -0.292 \log n_\mathrm{e} + 0.428 \left( \log \mathrm{sSFR} +9 \right). \label{eq:U(Z,ne,sSFR)}
\end{alignat}
All these coefficients are well constrained, fully rejecting zero coefficients, i.e., the null hypothesis of no dependence on the parameter.  In particular, we highlight that the scaling relation of $U (Z, n_\mathrm{e}, \mathrm{sSFR})$ is fairly constrained even with three variables.  The signs of these coefficients are all consistent with our initial assumption described in Section \ref{sec:ansatze} (see Figure \ref{fig:schem}).  Note that the magnitude of the coefficients does not immediately indicate the relative strength of the dependence on the corresponding parameter since the entire range covered by the parameters are different from each other.  

\begin{figure}
	\includegraphics[width=3.5in]{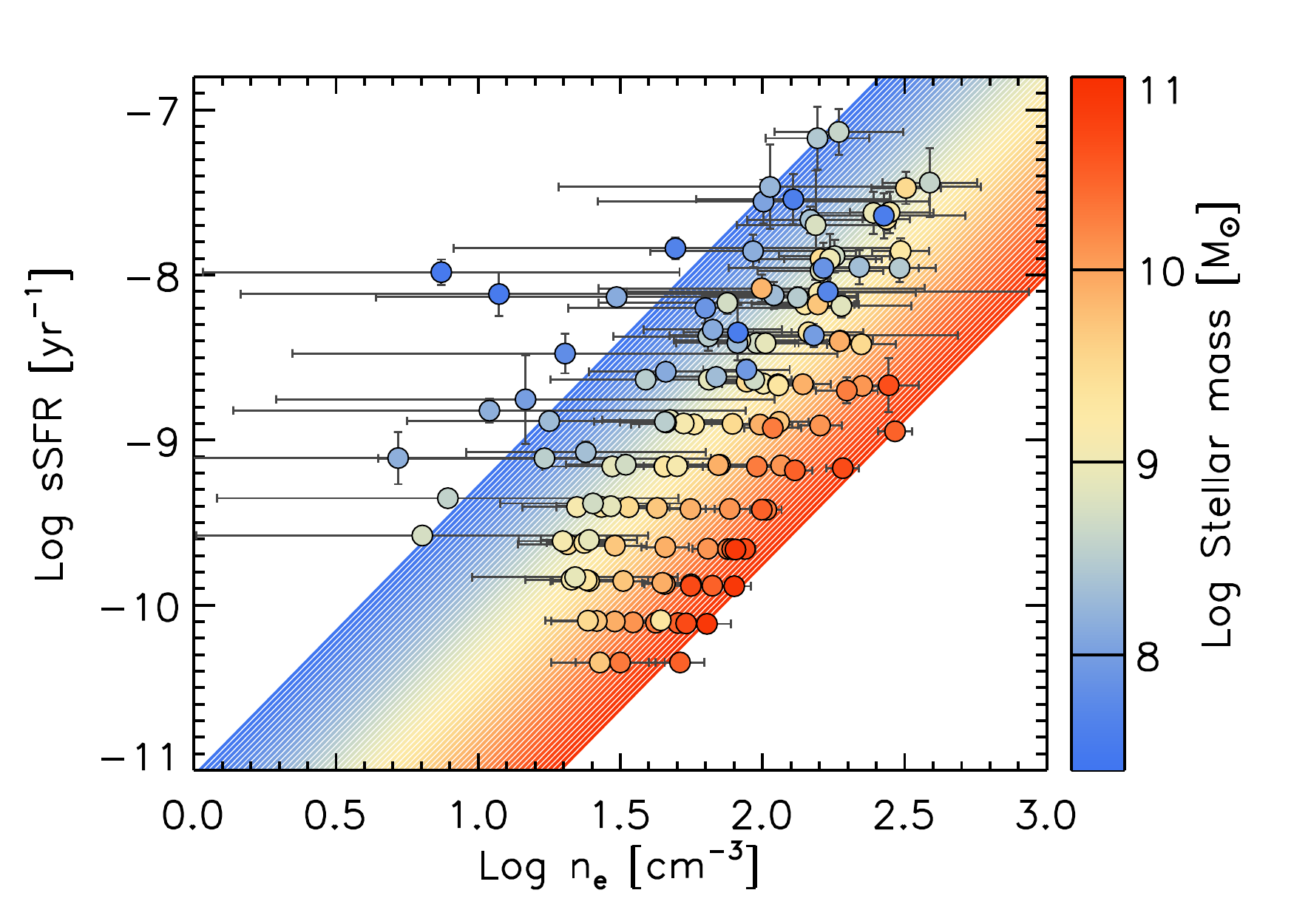}
	\caption{Specific SFR as a function of $n_\mathrm{e}$, color-coded by $M_\ast$.  The best-fit relation expressed by Equation \ref{eq:sSFR(M,ne)} is shown in the background.}
    \label{fig:n_e_vs_sSFR_2d_mass}
\end{figure}

\begin{figure} 
	\includegraphics[width=3.5in]{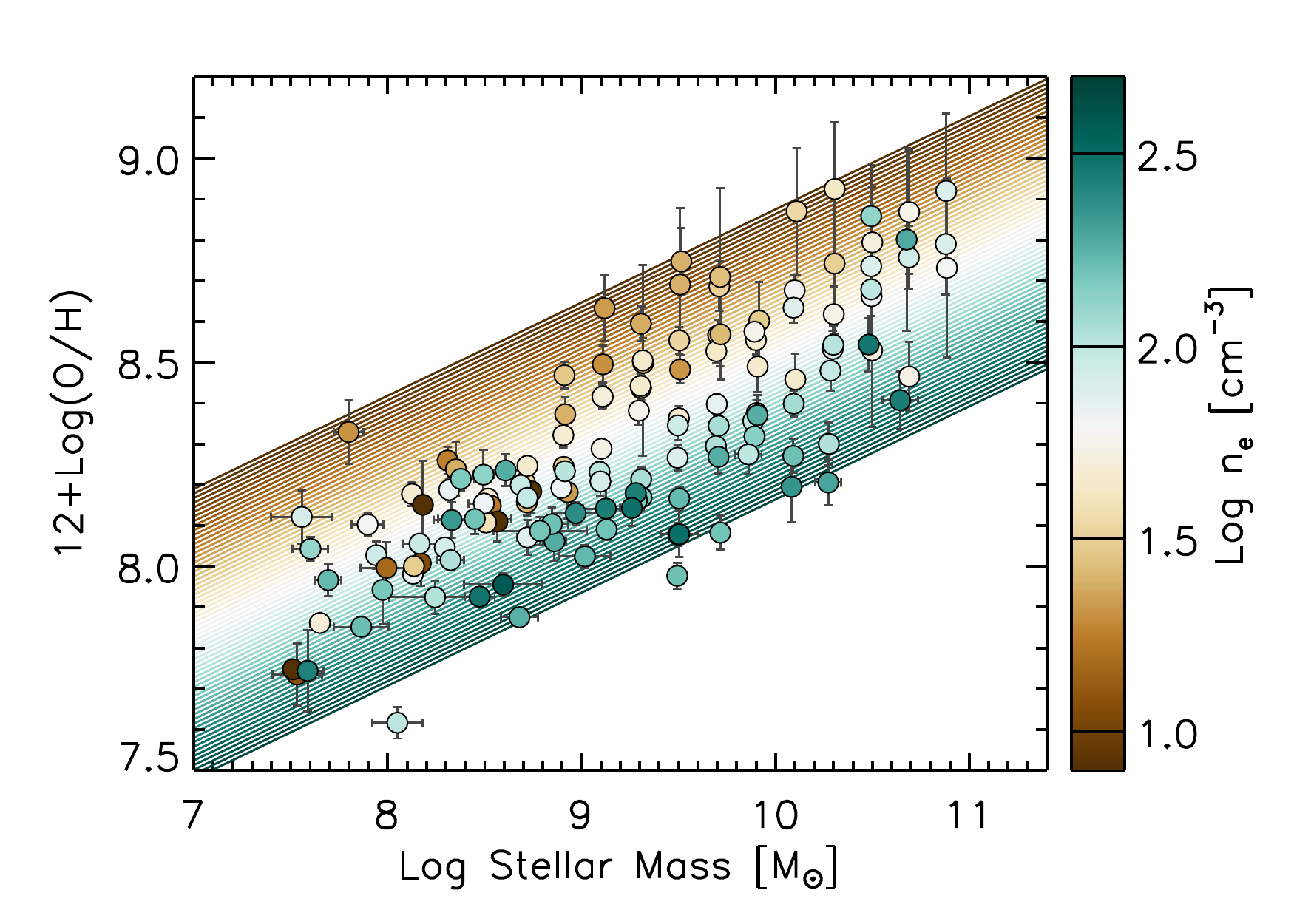}
	\includegraphics[width=3.5in]{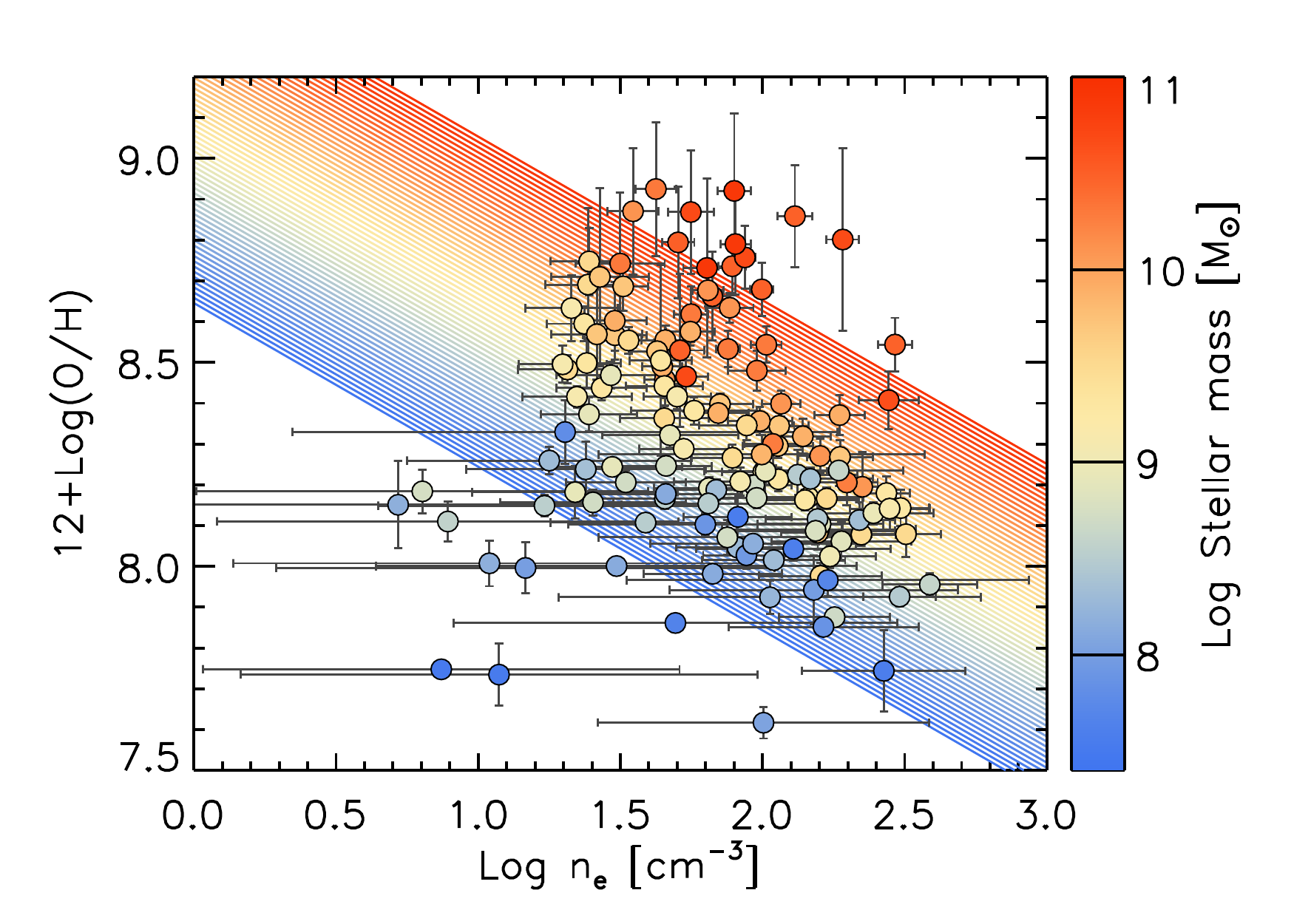}
	\caption{{\it Upper panel:} Metallicity $Z$ as a function of $M_\ast$, color-coded by $n_\mathrm{e}$.  {\it Lower panel:} $Z$ as a function of $n_\mathrm{e}$, color-coded by $M_\ast$.  The best-fit relation expressed by Equation \ref{eq:Z(M,ne)} is shown in the background.}
    \label{fig:n_e_vs_OH_2d_mass.eps}
\end{figure}

\begin{figure} 
	\includegraphics[width=3.5in]{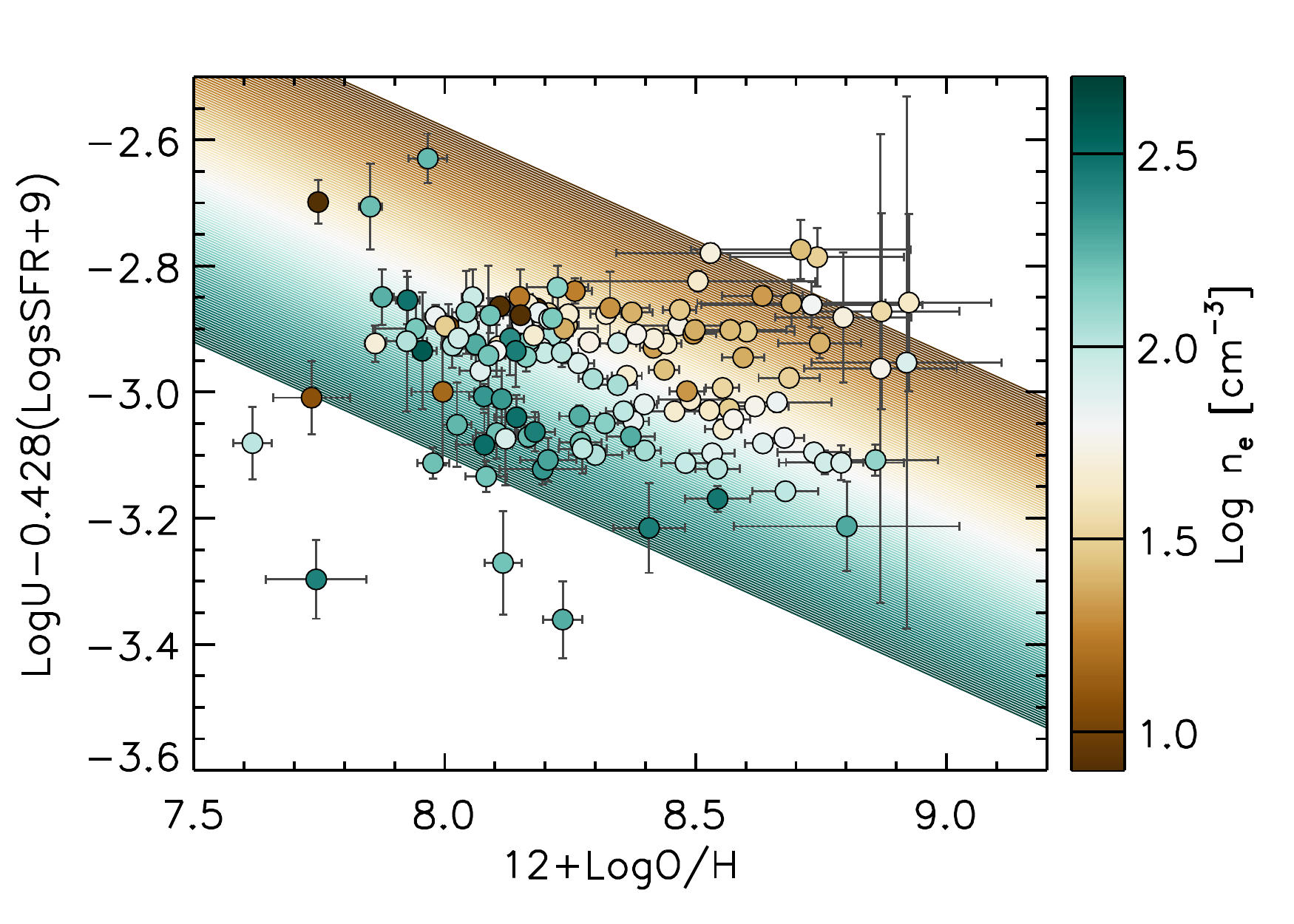}
	\includegraphics[width=3.5in]{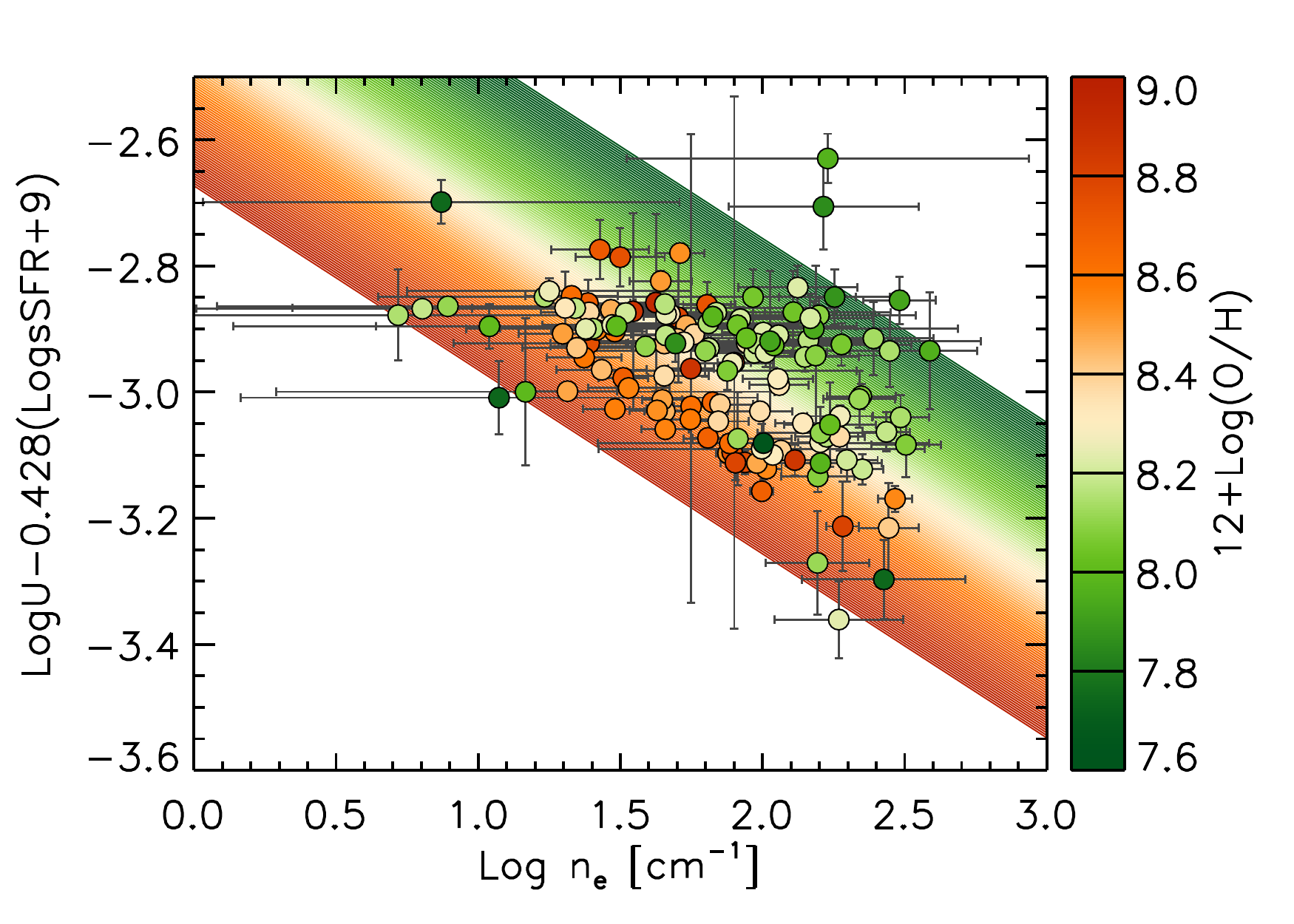}
	\caption{{\it Upper panel:} Residual of $U$ after subtracting the sSFR dependence ($0.428 \left( \log \mathrm{sSFR}+9 \right)$) as a function of $Z$, color-coded by $n_\mathrm{e}$.  {\it Lower panel:} $U - 0.428 \log \mathrm{sSFR}$ as a function of $n_\mathrm{e}$, color-coded by $Z$.  The best-fit relation expressed by Equation \ref{eq:U(Z,ne,sSFR)} is shown in the background.}
    \label{fig:dU_bsSFR3}
\end{figure}

In Figure \ref{fig:n_e_vs_sSFR_2d_mass}, we show the sSFR as a function of $n_\mathrm{e}$, color-coded by $M_\ast$, in comparison with the derived liner relation of Equation (\ref{eq:sSFR(M,ne)}) shown in the background.  Similarly, Figure \ref{fig:n_e_vs_OH_2d_mass.eps} shows $Z$ as a function of $M_\ast$, color-coded by $n_\mathrm{e}$ (upper panel), and as a function of $n_\mathrm{e}$, color-coded by $M_\ast$ (lower panel).  For $U$, we computed the residuals after subtracting the sSFR term ($0.428 \left( \log \mathrm{sSFR}+9 \right)$).  Figure \ref{fig:dU_bsSFR3} shows these residuals as a function of $Z$, color-coded by $n_\mathrm{e}$ (upper panel), and as a function of $n_\mathrm{e}$, color-coded by $Z$ (lower panel).  In all these panels of Figures \ref{fig:n_e_vs_sSFR_2d_mass}--\ref{fig:dU_bsSFR3}, the overall trends both of the correlation between the $x$- and $y$-axis variables and the color trend of the third axis variables are reasonably reproduced by the best-fit scaling relations.  Meanwhile, there are several outliers possibly due to their large errors on the estimates of $n_\mathrm{e}$ and/or $Z$.  In the next section, we attempt to interpret the global trends and discuss the physical mechanisms behind the scaling relations, not caring about minor deviations from these power-law relations.

\section{Discussions}
\label{sec:discussions}

We successfully constrained the dependencies of the physical quantities, $\mathrm{sSFR} (M_\ast, n_\mathrm{e})$, $Z (M_\ast, n_\mathrm{e})$, and $U (\mathrm{sSFR}, Z, n_\mathrm{e})$, based on the simplified ans{\"a}tze.   In this section, we attempt to interpret the results with our physically-motivated assumptions, and compare our findings to other theoretical predictions and observational results in the literature.

\subsection{Specific SFR and gas content}

Our result from regressing sSFR on $M_\ast$ and $n_\mathrm{e}$ indicates that sSFR is a strong function of $n_\mathrm{e}$ with a power-law slope of $\approx1.75$ at fixed $M_\ast$.  As noted above, this relatively large dependence includes the underlying correlations between $n_\mathrm{e}$ and $\mu_\mathrm{gas}$.  The sSFR of galaxies can be expressed in terms of SFE ($=\mathrm{SFR}/M_\mathrm{gas}$) and the gas content as follows:
\begin{equation}
\mathrm{sSFR} = \mathrm{SFE} \cdot M_\mathrm{gas} / M_\ast = \mathrm{SFE} \cdot \mu_\mathrm{gas}.
\label{eq:sSFR=SFEMgas/Ms}
\end{equation}
As we mentioned in Section \ref{sec:ansatze}, it would be natural to assume that the SFE is proportional to the inverse of the free-fall timescale $\tau_\mathrm{ff}$ of gas clouds, thus we assume $\mathrm{SFE} \propto n_\mathrm{e}^{1/2}$.  This simple assumption and the observed scaling of sSFR (Equation \ref{eq:sSFR(M,ne)}) yield
\begin{equation}
\mathrm{SFE} \propto \mathrm{sSFR}^{0.29} M_\ast^{0.18}.
\end{equation}
Observationally, \citet{2014ApJ...793...19S} found that SFE increases as a weak function of sSFR: $\mathrm{SFE}\propto \mathrm{sSFR}^{0.19}$, which may be in broad agreement with our result.  We can also obtain scaling relations involving the gas content as follows.
\begin{eqnarray}
\mu_\mathrm{gas} \propto M_\ast^{-0.63} n_\mathrm{e}^{1.3} \quad \textrm{or} \quad M_\mathrm{gas} \propto M_\ast^{0.37} n_\mathrm{e}^{1.3} \\
\Longleftrightarrow n_\mathrm{e} \propto \mu_\mathrm{gas}^{0.80} M_\ast^{0.50} \propto M_\mathrm{gas}^{0.80} M_\ast^{-0.30} \label{eq:n_e(Mgas,Ms)}.
\label{eq:ne_mugas}
\end{eqnarray}
This indicates that the electron density $n_\mathrm{e}$ in H\,{\sc ii} regions sensitively increases with the gas content at fixed $M_\ast$. We also obtain a relation between SFR and $M_\mathrm{gas}$ as 
\begin{equation}
\mathrm{SFR} \propto M_\mathrm{gas}^{1.40} M_\ast^{-0.15}.
\end{equation}

The scaling relation $\mathrm{SFR} \propto M_\mathrm{gas}^{1.40}$ (with minor dependence on $M_\ast$) agrees very well with observed integrated SK relation based on direct measurements of the gas content of galaxies \citep[e.g.,][]{2010ApJ...714L.118D,2010MNRAS.407.2091G,2013ApJ...768...74T,2014A&A...562A..30S}, as well as the conventional SK relation that correlates the surface densities of SFR and gas mass, for which observations have measured indices $N\approx1.4$ \citep[e.g.,][]{1998ApJ...498..541K}.  It is worth emphasizing that the scaling relation between sSFR and $\mu_\mathrm{gas}$ with a slope slightly smaller than 1.5, as supported by other observations, is derived by assuming $\mathrm{SFE}\propto \tau_\mathrm{tt}^{-1} \propto n_\mathrm{e}^{1/2}$.  In doing so, $n_\mathrm{e}$ is regarded as being proportional to the density of the molecular clouds where stars form, though this is not trivial because $n_\mathrm{e}$ is the measure of the electron density of H\,{\sc ii} regions.  On the other hand, the agreement between our result ($\mathrm{SFR} \propto M_\mathrm{gas}^{1.40}$) and other observations of the SK relations indicate that this assumption would be reasonable.

\subsection{Metallicity}
\label{sec:metallicity}

We found positive and negative dependences of metallicity $Z$, respectively, on $M_\ast$ and $n_\mathrm{e}$.  In the previous subsection, we obtained the scaling relation between $n_\mathrm{e}$ and gas content (Equation \ref{eq:ne_mugas}) based on a simple assumption of $\mathrm{SFE} \propto t_\mathrm{ff}^{-1} \propto n_\mathrm{e}^{1/2}$ and the observed scaling of sSFR (Equation \ref{eq:sSFR(M,ne)}).  Using Equation \ref{eq:ne_mugas}, we can eliminate $n_\mathrm{e}$ from the $Z(M_\ast, n_\mathrm{e})$ relation (Equation \ref{eq:Z(M,ne)}), and rewrite $Z$ in terms of $M_\ast$ and gas content as follows
\begin{equation}
Z \propto M_\ast^{0.23} n_\mathrm{e}^{-0.40} \propto \mu_\mathrm{gas}^{-0.32} M_\ast^{0.027}, 
\end{equation}
or, more simply, 
\begin{equation}
Z \propto M_\mathrm{gas}^{-0.32} M_\ast^{0.35}.
\end{equation}
The former equation indicates that the correlation between $Z$ and gas-to-stellar mass ratio $\mu_\mathrm{gas}$ would be almost independent of $M_\ast$ (and other parameters).  This is indeed what we expect from simple, classical models of chemical evolution of galaxies \citep[e.g.,][]{1963ApJ...137..758S}.  The latter form, however, indicates that the metallicity increases with $M_\ast$, as naturally expected as the result of the chemical evolution along the galaxy mass assembly history.  Meanwhile, the anticorrelation with gas content is as expected from the ``dilution'' scenario as described in Section \ref{sec:ansatze}.

Anticorrelations between $Z$ and the gas content of galaxies have been directly measured by observations.  \citet{2013MNRAS.433.1425B} analyzed a galaxy sample with the measurement of the H\,{\sc i} gas content, and found that the metallicity (via the strong-line method) decreases with the H\,{\sc i} gas mass $M_\mathrm{HI}$ at fixed $M_\ast$, approximately as $Z\propto M_\mathrm{HI}^{-0.17}$ (see Figure 2 of the reference).  Furthermore, \citet{2013MNRAS.434.2645D} attempted to model the global properties of the H\,{\sc i} gas content in galaxies using cosmological simulations.  Their preferable outflow model, which is able to reasonably reproduce the observed H\,{\sc i} mass function, predicts that the metallicity is inversely correlated with the H\,{\sc i} gas mass with a slope of $-0.26$ at a given $M_\ast$.  Our result is in reasonable agreement with theses dependencies found by both observations and numerical simulations.

The inverse correlation between $Z$ and sSFR (at fixed $M_\ast$) has been also characterized by regressing $Z$ on the $M_\ast$--sSFR plane (Table \ref{tb:coeff} lower part) as follows:
\begin{equation}
12+\log (\mathrm{O/H}) = 8.427 + 0.136 \left(\log M_\ast -10 \right) -0.208 \left(\log \mathrm{sSFR}+9\right).
\label{eq:Z(M,sSFR)}
\end{equation}
We note that this relation is nearly equivalent to what is obtained by eliminating $n_\mathrm{e}$ from $Z(M_\ast, n_\mathrm{e})$ and $n_\mathrm{e} (M_\ast, \mathrm{sSFR})$ (see Table \ref{tb:coeff}), with a slight difference in the coefficient of $\log M_\ast$.  The relation between these three parameters ($M_\ast$, $Z$, sSFR), i.e., the FMR, has been commonly characterized by introducing a variable $\mu_\alpha$ defined as follows:
\begin{equation}
\mu_\alpha = \log M_\ast - \alpha \log \mathrm{SFR}.
\end{equation}
where the parameter $\alpha$ is determined to minimize the scatter in terms of metallicity around the average relation.  Our result from a simple plane fitting provides $\alpha = 0.208/(0.136+0.208) = 0.60\pm 0.01$.  This is in good agreement with the value ($\alpha=0.66$) found by \citet{2013ApJ...765..140A}.  Meanwhile, such values of $\alpha$ are larger than those ($\alpha \approx 0.3$) as reported by studies relying on the strong-line metallicity indicators \citep[e.g.,][]{2010MNRAS.408.2115M}.  Note that Equation (\ref{eq:Z(M,sSFR)}) can be rewritten using $\mu_{0.61}$ as follows: 
\begin{equation}
12 + \log (\mathrm{O/H})  = (5.19\pm0.06) + (0.344\pm0.006)\mu_{0.60}.
\end{equation}
The slope of $0.34$ is lower than the slope of 0.43 found by \citet{2013ApJ...765..140A}, and the difference is not distinguished if we adopt $\alpha=0.66$.

\subsection{Ionization parameter}

Our results indicate that the ionization parameter $U$ varies as a strong function of the sSFR of galaxies, but also depends significantly on both $Z$ and $n_\mathrm{e}$.  The former dependence implies that the ionization photon flux in a single H\,{\sc ii} region increases on average with increasing {\it total} (i.e., galaxy-wide) sSFR of the host galaxies.  To directly examine this hypothesis, we need to resolve the individual H\,{\sc ii} regions in galaxies across wide ranges of $M_\ast$ and sSFR.

There are other ideas to account for the correlations between sSFR and $U$.  The observed [O\,{\sc iii}]/[O\,{\sc ii}] ratio, a proxy of $U$, could effectively increase for a ``density-bounded'' H\,{\sc ii} region, which is fully ionized and whose size is determined by the cloud size, instead of the Str{\"o}mgren radius.  Since O$^{++}$ dominates at the inner region of the ionized sphere, close to the radiation source, whereas O$^{+}$ becomes more dominant at the outer region, density-bounded H\,{\sc ii} regions are expected to have a significantly smaller [O\,{\sc ii}] zone, resulting in higher [O\,{\sc iii}]/[O\,{\sc ii}].  Specially, \citet{1996A&A...312..365B} considered the effects of density-bounded (``matter-bounded'' in the reference) ionized gas clouds to the changes of the emission-line ratios, and showed that the line ratios (especially, [O\,{\sc iii}]/[O\,{\sc ii}]) vary along the increasing fraction of the density-bounded clouds in a very similar way to the variation with increasing $U$ (see Figure 6 in the reference)\footnote{This study specially focused on the conditions in the narrow-line region of active galactic nuclei, but can also applied to star-forming H\,{\sc ii} regions.}.  A larger contribution of density-bounded H\,{\sc ii} regions has proposed to account for observed higher [O\,{\sc iii}]/[O\,{\sc ii}] ratios \citep{2008MNRAS.385..769B,2014MNRAS.442..900N}.  In addition, it is recently thought that extremely high [O\,{\sc iii}]/[O\,{\sc ii}] ratios are often incident to non zero escape fraction, i.e., the fraction of ionizing photons that escape into intergalactic space \citep[e.g.,][]{2014MNRAS.442..900N,2016ApJ...831L...9N,2016MNRAS.461.3683I,2018MNRAS.478.4851I}.  Similarly, if an H\,{\sc ii} region overlaps an adjacent H\,{\sc ii} region, the effective ionization parameter would increase locally because the overlap region is illuminated by both ionizing sources.  It could be naturally expected that such situations tend to happen more frequently in systems with higher sSFRs because the number density of H\,{\sc ii} regions may increase with increasing sSFR.  We note that our estimate of the ionization parameter assumes a simple ionization-bounded plane-parallel geometry in order to limit the number of free parameters, and that our estimation could thus be inaccurate for some cases.   We defer further investigations of the relations between $U$ and the emission-line indicators, such as [O\,{\sc iii}]/[O\,{\sc ii}], including more realistic geometric conditions to future work.

We now turn to the dependence on metallicity.  There have been observational studies that indicate the presence of an anticorrelation between $U$ and $Z$ based on a strong-line method \citep{2014MNRAS.442..900N,2016ApJ...822...42O}.  \citet{2015ApJ...801...88S} and \citet{2017PASJ...69...44K} mentioned the anticorrelation by using the stacked measurements based on the direct method from \citet{2013ApJ...765..140A}.  In Figure \ref{fig:triangle}, our data show that $U$ is apparently strongly anticorrelated with $Z$: a simple power-law fit yields $U \propto Z^{-1.52\pm0.04}$.  Compared with this {\it apparent} relation, however, our result from the four-dimensional fitting indicates a shallower slope for the {\it partial} correlation ($U \propto Z^{-0.36}$; Equation \ref{eq:U(Z,ne,sSFR)}) at fixed sSFR and $n_\mathrm{e}$.  The apparent stronger $Z$-dependence can be attributed to the anticorrelation between sSFR and $Z$ (Section \ref{sec:metallicity}).  Our result thus indicates necessity of physical mechanisms to generate the intrinsic {\it moderate} inverse correlation between $U$ and $Z$.  Some explanations have been suggested to account for the $U$-$Z$ anticorrelation as mentioned in Section \ref{sec:ansatze} \citep[e.g.,][]{1986ApJ...307..431D,2006ApJS..167..177D,2013ApJ...774..100K}.

\citet{2006ApJ...647..244D} have theoretically found that $U$ correlates with $Z$ as $U \propto Z^{-0.8}$ including the following two effects.  First, at higher metallicity, ionizing photons emitted by a star are more absorbed by the stellar winds due to an increased optical depth.  Second, at higher metallicity, ionizing photons are more efficiently scattered in the photospheres and their radiation energy is more converted to kinetic energy in the stellar wind base region \citep{1992ApJ...401..596L}.  It may be worth noting that the line-blanketing effects could also be considered as an origin of the $U$--$Z$ anticorrelation: metal elements stored in the stellar photosphere block the radiated ionizing photons, reducing the emergent flux at wavelengths of the metallic lines.  However, the back-warming effect occurs immediately following the blocking, where metals re-emit photons at lower frequencies, raising the adjacent continuum.  \citet{2002ApJ...577..389K} demonstrated that these effects have little influence on the number of hydrogen ionizing photons ($\ge 13.6~\mathrm{eV}$) because the two effects almost entirely cancel each other.  

Another explanation is that the $U$--$Z$ correlation is in part caused by dust absorption in the H\,{\sc ii} region.  \citet{2001ApJ...555..613I} established a theoretical framework for the dust absorption of ionizing photon in an H\,{\sc ii} region, and found observationally an anticorrelation between the fraction $f_\gamma$ of the ionizing photons contributing to ionization and metallicity (or dust-to-gas mass ratio) for individual H\,{\sc ii} regions in the Milkey Way.  Similarly, \citet{2001AJ....122.1788I} found signs of such anticorrelation also for nearby galaxies, with $f_\gamma$ ranging from $\sim40\textrm{--}70\%$.

The $U$--$Z$ anticorrelation could also be related to the possible correlation between the IMF and metallicity \citep{1985ApJS...58..125E}: a top-heavy IMF would yield more ionizing photons per SFR.  Recently, \citet{2015ApJ...806L..31M} found that the local IMF is tightly related to the local metallicity, becoming top-heavier at lower metallicity.  Such correlation may be explained if the fragmentation scales of the molecular clouds are determined by the Jeans mass, which increases with increasing temperature (i.e., towards lower metallicity).  A change in the IMF, however, may also change the estimations of $M_\ast$ and SFR, as well as the ionization parameter because the shape of the ionizing spectrum changes.  We defer further investigation including the possible dependence of the IMF to future work.

Turning to the $n_\mathrm{e}$-dependence, our four-dimensional fitting found a negative {\it partial} correlation between $U$ and $n_\mathrm{e}$ as $U\propto n_\mathrm{e}^{-0.29}$, at fixed sSFR and $Z$, as opposed to their {\it apparent} positive correlation (with the Spearman's rank correlation coefficient $\rho = 0.476$ with $p$-value $<10^{-8}$) as shown in the relevant panel in Figure \ref{fig:triangle}.   The apparent correlation is caused by the positive correlation between sSFR and $n_\mathrm{e}$ and the tight correlation between $U$ and sSFR.  Indeed, the bottom-leftmost panel in Figure \ref{fig:triangle} shows the presence of an inverse correlation between $U$ and $n_\mathrm{e}$ at fixed sSFR (i.e., at fixed color).  The observed dependence is weaker than what is expected from the definition of $U$, i.e., $U \propto n_\mathrm{e}^{-1}$.  This may indicates that the differential electron densities between relatively inner regions, where the ionization parameter is defined, and outer regions, which may dominate the [S\,{\sc ii}] fluxes, in the H\,{\sc ii} regions.  

The weak negative dependence of $U$ on $n_\mathrm{e}$ is in good agreement with what was derived by \citet{2006ApJ...647..244D}, who found that the theoretical ionization parameter is weakly related to the density $n_0$ of the ambient ISM as $ U\propto n_0^{-1/5}$ by modeling the evolution of an expanding H\,{\sc ii} region, pressurized by the stellar winds from the central star cluster.  In their model, the internal density $n_\mathrm{in}$ of an isothermal spherical H\,{\sc ii} region is related to the ambient density, as $n_\mathrm{in} \propto n_0^{3/5}$, as well as the inner radius $R_\mathrm{in}$ is also linked to it as $R_\mathrm{in} \propto n_0^{-1/5}$ \citep{1997MNRAS.289..570O}.   For a fixed ionizing luminosity of the central source, the ionization parameter depends on these quantities as $U\propto 1/(n_\mathrm{in}R_\mathrm{in}^2)$, yielding the relation between $U$ and the ambient density as above.

Lastly, we note that we recently obtained a conclusion that the ionization parameter $U$ is directly linked to sSFR, while little or probably not with metallicity \citep{2018MNRAS.tmp..962K}, from comparison between the high-redshift sample of the main-sequence star-forming galaxies at $z\sim 1.6$ (a subset from \citealt{2017ApJ...835...88K,2018arXiv181201529K}) and the ``sSFR-matched'' sample of  the SDSS galaxies.  These two samples that have the equivalent  average sSFRs present similar levels of the ionization parameter, whereas the sSFR-matched low-redshift sample shows relatively lower metallicities.  However, our results in this paper are indeed not against what is found by \citet{2018MNRAS.tmp..962K} within the uncertainties.  The difference in the metallicity of these two samples is found to be $\sim 0.1~\mathrm{dex}$.  From our result, this metallicity difference induces only a very small difference ($\lesssim 0.05~\mathrm{dex}$) in $U$ at fixed sSFR, which is equivalent to the level of statistical uncertainties presented in \citet{2018MNRAS.tmp..962K}.  We also note that our analysis has been applied only for the low redshift galaxies, whereas the attention in \citet{2018MNRAS.tmp..962K} was more focused on the evolution over a wide range of cosmic time.  The investigation of redshift evolution of the relationships that we found between the nebular parameters is beyond the scope of this paper.

\subsection{Application of the scaling relations}

Lastly we mention possible application of the scaling relations we derived in this paper.  Our results can be implemented into a subgrid recipe of galaxy evolution simulations to compute the conditions of ionized nebular from the properties of host galaxies, and predict the fluxes of emission lines of interest.  Predicting strong-line fluxes is useful to prepare mock catalogs for future spectroscopic surveys as well as to test the models by comparing observations.  Our results would be also implemented into SED fitting codes to include the contribution from emission lines for given galaxy properties.

Here we present an application of our scaling relations for a semi-analytic model galaxy catalog.  In semi-analytic models, the evolution of galaxies embedded in collapsed dark halos, is traced following the halo merger trees while solving a simplified model that controls the star formation involving relevant physical phenomena such as supernova and/or AGN feedback.  As a result, some fundamental quantities of the galaxies, such as $M_\ast$, $M_\mathrm{gas}$, SFR, metal abundance and the stellar population, are obtained as functions of cosmic time.  Once $M_\ast$ and $M_\mathrm{gas}$ are given, our scaling relations can be utilized to compute the conditions of the H\,{\sc ii} regions, $n_\mathrm{e}$ and $U$, from Equations (\ref{eq:n_e(Mgas,Ms)}) and (\ref{eq:U(Z,ne,sSFR)}), respectively.  With the gas-phase $Z$ traced in the simulation, it is then possible to compute the emission line fluxes by interpolating a prepared table of the line fluxes for grids of ($n_\mathrm{e}, Z, U$).  

\begin{figure*}
	\includegraphics[width=5.1in]{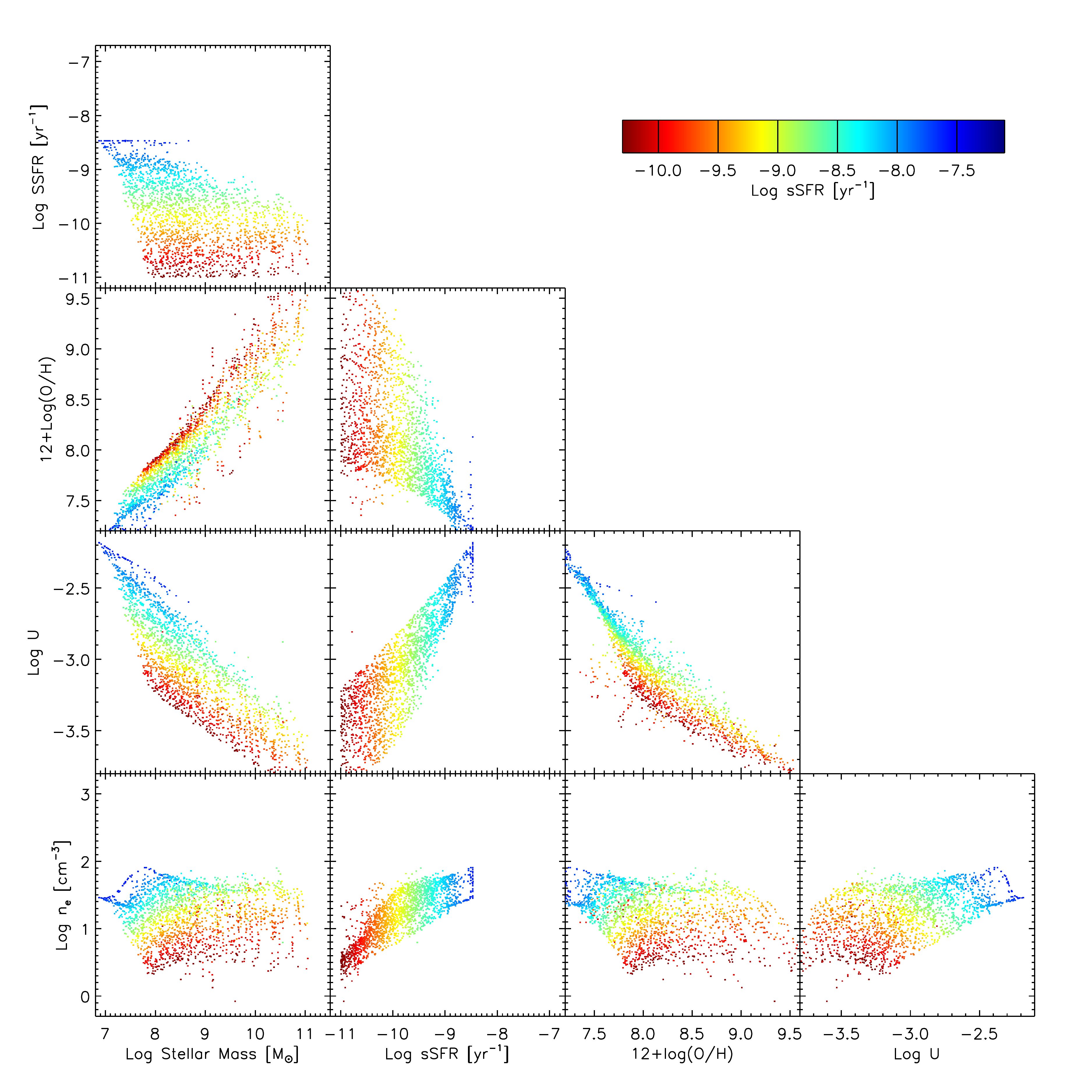}
	\caption{Application of the derived scaling relations to a semi-analytic model galaxy sample.  Same as Figure \ref{fig:triangle}, but for 4855 simulated galaxies taken from the $\nu^2$GC light-cone ($\mathrm{redshift}<0.3$, $\mathrm{sSFR}\ge10^{-11}~\mathrm{yr^{-1}}$).  The values of $n_\mathrm{e}$ and $U$ are computed using Equations (\ref{eq:n_e(Mgas,Ms)}) and (\ref{eq:U(Z,ne,sSFR)}) from $M_\ast$, sSFR, and $Z$ given by the simulation.  The sSFR range of the color scale is different from Figure \ref{fig:triangle}.}
    \label{fig:triangle_n2gc}
\end{figure*}

For a demonstration, we use the New Numerical Galaxy Catalog ($\nu^2$GC; \citealt{2015PASJ...67...61I,2016PASJ...68...25M})\footnote{http://hpc.imit.chiba-u.jp/~nngc/index.html}.  From the public light-cone catalog, we extracted a sample of 4855 galaxies limiting to those at $z<0.3$ and having $\mathrm{sSFR}\ge10^{-11}~\mathrm{yr^{-1}}$.  We define the stellar mass as the sum of disk ({\tt Mstard}) and bulge ({\tt Mstarb}) masses, and take the gas mass ($M_\mathrm{gas}=${\tt Mcool}) and metal mass ($M_Z=${\tt MZc}) in the cold phase.  The metallicity is defined as $12+\log (\mathrm{O/H}) = 8.69 + \log (M_Z/M_\mathrm{gas}/Z_\odot)$ where $Z_\odot=0.014$.  In Figure \ref{fig:triangle_n2gc}, we show the correlations among $M_\ast$, sSFR, $Z$, given by the semi-analytic model, and $n_\mathrm{e}$ and $U$ computed with our scaling relations.  In the $\nu^2$GC sample, galaxies with $\mathrm{sSFR}\ge10^{-8.5}~\mathrm{yr^{-1}}$ are missing, while a space of low $M_\ast$ ($<10^9~M_\odot$) and low sSFR ($<10^{-9.5}~\mathrm{yr^{-1}}$) is filled up, where we have no SDSS stacks (see Figure \ref{fig:triangle}).  Moreover, a population of very metal-rich galaxies ($12+\log\mathrm{O/H}>9$) exits.  Regardless of these discrepancies, the trends of the SDSS sample seen in Figure \ref{fig:triangle} are globally reproduced here, manifesting the effectiveness of the application to semi-analytic model galaxy samples.  Detailed comparison of the $\nu^2$GC catalog with observations and/or predictions of any concrete observables are beyond the scope of this paper.

\section{Conclusions}
\label{sec:conclusions}

We have investigated the connection between the physical parameters characterizing the conditions of star-forming ionized gas and the global properties of the host galaxies, especially focusing on five key parameters, i.e., stellar mass, sSFR, oxygen abundance, ionization parameter, and electron density of the ionized gas.  In order to obtain accurate estimates of the oxygen abundance, we have utilized the direct method based on the detection of auroral lines ([O\,{\sc iii}]$\lambda$4363, and [O\,{\sc ii}]$\lambda\lambda$7320,7330) in the set of stacked spectra of the SDSS galaxies at $z\sim0.1$, which are binned into the $M_\ast$--sSFR grids.  Our key conclusions can be summarized as follows.
\begin{enumerate}
\item The gas-phase oxygen abundance was successfully measured for a wide range of $M_\ast$ and sSFR by using stacked spectra.  The resultant values range between $7.6\lesssim 12+\log (\mathrm{O/H}) \lesssim 8.9$, tightly correlating with sSFR and $M_\ast$.
\item The electron density $n_\mathrm{e}$ was measured from the [S\,{\sc ii}] doublet ratio, accounting for the dependence on the electron temperature, and found to increase with increasing sSFR at fixed $M_\ast$ as well as with increasing $M_\ast$ at fixed sSFR.   
\item The ionization parameter $U$ was measured from the [O\,{\sc iii}]/[O\,{\sc ii}] ratio, accounting for the dependencies on $Z$ and $n_\mathrm{e}$.  The resultant values of $U$ were found to be tightly correlated with the sSFR, and also to depend on $Z$ and $n_\mathrm{e}$.
\item The partial anticorrelation between $U$ and $Z$ ($U \propto Z^{-0.36}$ at fixed sSFR and $n_\mathrm{e}$) was found to be weaker than the {\it apparent} correlation ($U \propto Z^{-1.52}$).  The observed moderate (i.e., not as strong as its apparent correlation) anticorrelation between $U$ and $Z$ is qualitatively consistent with theoretical predictions of the intrinsic $U$--$Z$ relation.
\item The ionization parameter is also found to negatively correlate with $n_\mathrm{e}$ at fixed sSFR and $Z$ as $U\propto n_\mathrm{e}^{-0.29}$.  This is qualitatively consistent with the definition of the ionization parameter, though the dependence on $n_\mathrm{e}$ is weaker than the inverse proportion. This may indicates that the electron density probed by the [S\,{\sc ii}] doublet scales with the inner surface density more sensitively than proportionality.
\end{enumerate}

Our results presented in this paper have firmly provided the global picture on the behavior of the nebulae at the galaxy-wide scales.  The scaling relationships that we found between the fundamental properties of galaxies and the nebulae provide a set of boundary conditions which are to be reproduced by high-resolution simulations.  On the other hand, our results provide a convenient tool to compute the nebular conditions and to predict fluxes of strong emission lines with little computational cost in semi-analytic models and/or SED fitting codes.  Such application is thus useful to construct mock catalogs for forthcoming large spectroscopic surveys with the next-generation spectrographs, such as the the Subaru/Prime Focus Spectrograph (PFS), Dark Energy Spectroscopic Instrument (DESI), Multi-Object Optical Near-infrared Spectrograph (MOONS), and Euclid and WFIRST in space.  These next-generation surveys will then be enable us to extend our analysis beyond the low redshift Universe in the near future, and to understand the physical processes behind the connections of galaxy properties in more detail.

\section*{Acknowledgements}

We are grateful Brett Andrews for kindly providing us with their catalog and data.  We also thank K.~Yabe and T.~Kojima for useful discussions and comments.  This work has been supported by KAKENHI (17H01114) and by NAOJ ALMA Scientific Research Grant number 2016-01 A.

Funding for the SDSS and SDSS-II has been provided by the Alfred P. Sloan Foundation, the Participating Institutions, the National Science Foundation, the U.S. Department of Energy, the National Aeronautics and Space Administration, the Japanese Monbukagakusho, the Max Planck Society, and the Higher Education Funding Council for England. The SDSS Web Site is http://www.sdss.org/.

The SDSS is managed by the Astrophysical Research Consortium for the Participating Institutions. The Participating Institutions are the American Museum of Natural History, Astrophysical Institute Potsdam, University of Basel, University of Cambridge, Case Western Reserve University, University of Chicago, Drexel University, Fermilab, the Institute for Advanced Study, the Japan Participation Group, Johns Hopkins University, the Joint Institute for Nuclear Astrophysics, the Kavli Institute for Particle Astrophysics and Cosmology, the Korean Scientist Group, the Chinese Academy of Sciences (LAMOST), Los Alamos National Laboratory, the Max-Planck-Institute for Astronomy (MPIA), the Max-Planck-Institute for Astrophysics (MPA), New Mexico State University, Ohio State University, University of Pittsburgh, University of Portsmouth, Princeton University, the United States Naval Observatory, and the University of Washington.




\bibliographystyle{mnras}
\bibliography{ads} 




\appendix

\section{Balmer absorption}
\label{sec:BalmerAbsorption}

Here we present empirically-calibrated formulae for correcting the observed fluxes of the Balmer lines (H$\alpha$ and H$\beta$) for the reduction due to the stellar atmospheric absorption as a function of $M_\ast$ and sSFR.  To evaluate the effects of the Balmer absorption, we compare the flux measurements on the composite spectra in which the stellar continuum is subtracted using pPXF (Section \ref{sec:stesub}) with those based on the continuum subtraction using a simple linear fitting, as usually done for continuum-faint or low-S/N spectra.  From these measurements, we defined the fraction of the absorbed flux $f_\mathrm{abs}$ for each stack, $f_\mathrm{abs} = (F_\mathrm{int}-F_\mathrm{obs})/F_\mathrm{int}$, where $F_\mathrm{obs}$ and $F_\mathrm{int}$ are the flux measured, respectively, with the linear-fit based and pPXF-based continuum subtraction.

\begin{figure*}
   \includegraphics[width=6in]{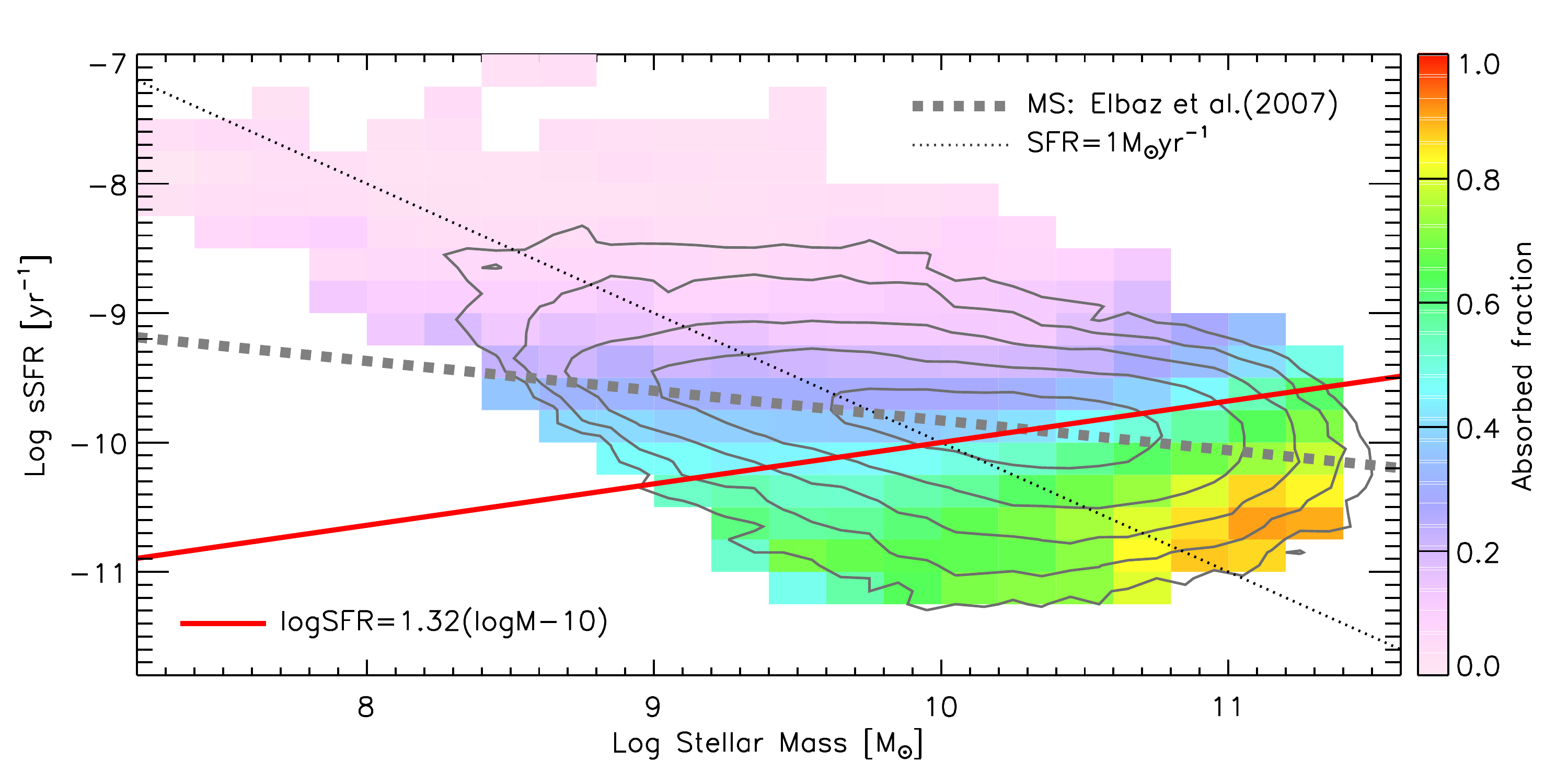}
   \caption{The fraction of the absorbed H$\beta$ flux relative to the intrinsic value is shown by colors in the $M_\ast$ vs. sSFR plane.  Each box represents a single stack.  For reference, we overplot the main-sequence of local galaxies (thick dotted line; \citealt{2007A&A...468...33E}) and a relation corresponding to $\mathrm{SFR}=1~M_\odot~\mathrm{yr^{-1}}$ (thin dotted line).  Red solid line indicates a relation expressed as $\log \mathrm{SFR}/(M_\odot~\mathrm{yr^{-1}}) = 1.32\times \left(\log M_\ast/M_\odot -10\right)$.}
   \label{fig:M_vs_SFR_HB_BalmerAbs}
\end{figure*}

\begin{figure}
   \includegraphics[width=3.5in]{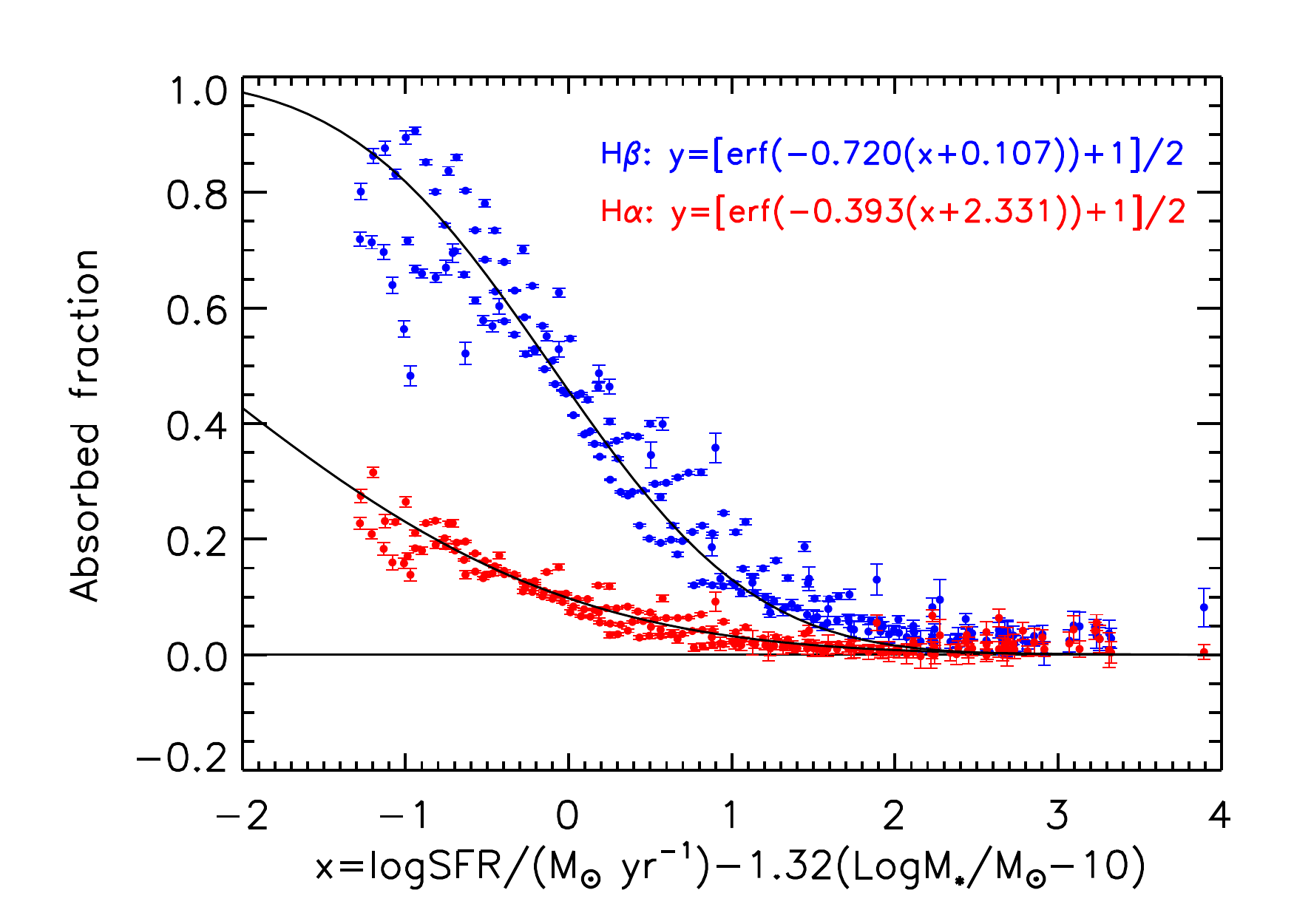}
   \caption{The fraction of the absorbed flux of H$\alpha$ (red) and H$\beta$ (blue) each as a function of $x = \log \mathrm{SFR}/(M_\odot~\mathrm{yr^{-1}}) -1.32 (\log M_\ast/M_\odot -10)$.  Solid curves indicate the best-fit empirical relations for each line  (Equations \ref{eq:balmerabs_Ha} and \ref{eq:balmerabs_Hb}).}
   \label{fig:x_fabs}
\end{figure}

In Figure \ref{fig:M_vs_SFR_HB_BalmerAbs}, we show the absorbed fraction of the H$\beta$ line in the $M_\ast$--sSFR plane.  Stacks are limited to those with $S/N(\mathrm{H\beta})\ge 40$.  Each box represents a single stack, color-coded by $f_\mathrm{abs}(\mathrm{H\beta})$.  It is clear that a strong trend exists between $f_\mathrm{abs}$ and a quantity, which can be defined as a combination of $M_\ast$ and SFR.  The absorption fraction of the H$\alpha$ flux exhibits a similar trend to H$\beta$, but the absolute values are much smaller.  Considering a linear combination as $x_\alpha = \log \mathrm{SFR} - \alpha \log M_\ast$, we found that ,when $\alpha=1.32$, the correlation becomes the tightest for both H$\alpha$ and H$\beta$.   The relation $x_\mathrm{\alpha=1.32} = 0$ is shown by a red solid line.  In Figure \ref{fig:x_fabs}, we show the absorption fraction of the H$\alpha$ and H$\beta$ fluxes as a function of $x_{\alpha=1.32}$.  To express these trends, we fit an empirical functional form to data, and obtained the forms as follows,
\begin{eqnarray}
f_\mathrm{abs} (\mathrm{H\alpha}) = \frac{1}{2}\left[ \mathrm{erf} \left(-0.373 \left(x + 2.331\right) \right) + 1\right], \label{eq:balmerabs_Ha} \\
f_\mathrm{abs} (\mathrm{H\beta}) = \frac{1}{2}\left[ \mathrm{erf} \left(-0.720 \left(x + 0.107\right) \right) + 1\right]. \label{eq:balmerabs_Hb}
\end{eqnarray}

\section{Nebular diagnostic diagrams}
\label{sec:diagnostic}

\begin{figure*}
	\includegraphics[height=9in]{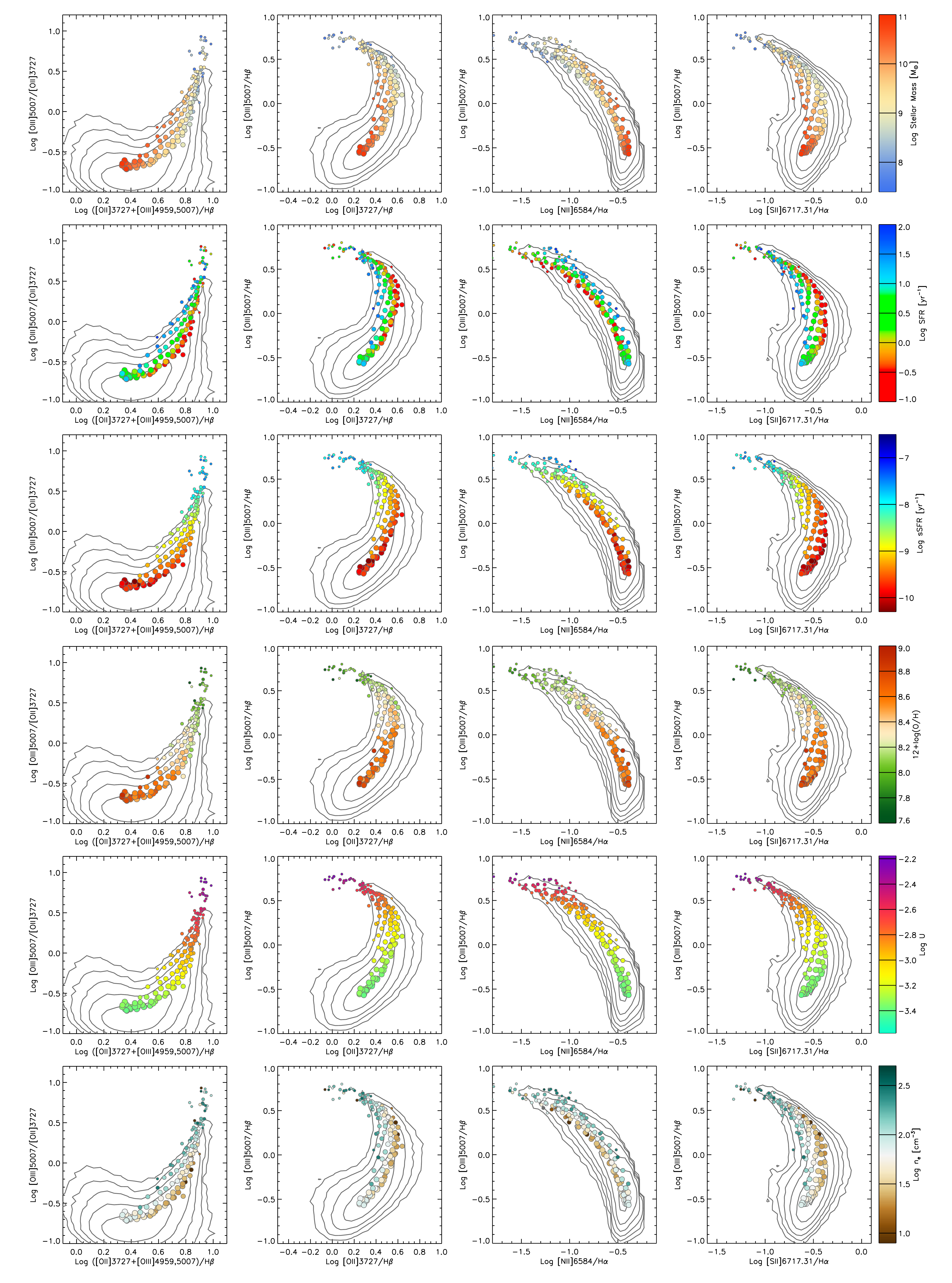}
	\caption{$R_\mathrm{23}$ vs. $O_\mathrm{32}$,  $O_2$ vs. $O_3$, $N_2$ vs. $O_3$, and $S_2$ vs. $O_3$ diagrams from left to right.  The SDSS stacks are shown by circles, color-coded by different properties of host galaxies ($M_\ast$, SFR, and sSFR; upper 12 panels) and nebular parameters ($Z$, $U$, $n_\mathrm{e}$; lower 12 panels).  Symbol size corresponds to the number of galaxies in the stack (see Figure \ref{fig:triangle}).  Contours show the distribution of the individual galaxies in the parent sample.}
    \label{fig:BPT}
\end{figure*}

It would be useful to see the relation between the strong-line flux ratios and the galactic and nebular properties.  In Figure \ref{fig:BPT}, we show four routinely-used diagrams -- $R_{23}$ vs. $O_{32}$, $O_2$ vs. $O_3$, $N_2$ vs. $O_3$, and $S_2$ vs. $O_3$ diagrams (the so-called BPT diagrams; \citealt{1987ApJS...63..295V}).  Here the line ratio indices are defined as follows: 
$R_{23} = ( \textrm{[O\,{\sc ii}]}\lambda3727 + \textrm{[O\,{\sc iii}]}\lambda\lambda4959,5007)/\mathrm{H\beta}$, 
$O_{32} = \textrm{[O\,{\sc iii}]}\lambda5007/\textrm{[O\,{\sc ii}]}\lambda3727$, 
$O_{2} = \textrm{[O\,{\sc ii}]}\lambda3727/\mathrm{H\beta}$, 
$O_{3} = \textrm{[O\,{\sc iii}]}\lambda5007/\mathrm{H\beta}$, 
$N_{2} = \textrm{[N\,{\sc ii}]}\lambda6584/\mathrm{H\alpha}$, 
and $S_{2} = \textrm{[S\,{\sc ii}]}\lambda\lambda6717,6731/\mathrm{H\alpha}$.  
The contours show the distribution of the individual SDSS galaxies in the parent catalog.  To draw the contours, we additionally impose on the sample the detection of [O\,{\sc iii}]$\lambda$5007 at $>3\sigma$ (89 percent of the sample), although the original selection does not care about the detection of [O\,{\sc iii}]$\lambda$5007 (Section \ref{sec:sample}).  The line ratios of the individual sources are corrected for extinction using the \citet{1989ApJ...345..245C} reddening curve and assuming the intrinsic H$\alpha$/H$\beta = 2.86$.   The SDSS stacks with the successful metallicity estimation are shown by circles, color-coded by $M_\ast$, SFR, sSFR, $12+\log(\mathrm{O/H})$, $U$, and $n_\mathrm{e}$ in each panel.  

We can see clear trends of the location in the diagrams against all the average physical properties considered here.  The trends with $M_\ast$ are consistent with those of the individual galaxies \citep[][see Figure 2]{2016ApJ...828...18M}.  At a fixed SFR, the stacks cover a wide range of the emission-line ratios, almost from the one to the other tail-end in all four diagrams (second row from top in Figure \ref{fig:BPT}), forming a narrow sequence parallel to the ridge line of the contours.  The nebular conditions are also tightly correlated with the location in the diagram.  The metallicity varies along the sequence, from the lowest $Z$ at the high end-tail of the $y$-axis values to the highest $Z$ at the opposite side (forth row in Figure \ref{fig:BPT}).  The trend with $U$ is very similar to sSFR (fifth row).  This is corresponding to the tight correlation between $U$ and sSFR which we have seen in Figure \ref{fig:triangle}.  The electron density $n_\mathrm{e}$ is correlated with the location in a similar way to SFR: stacks of fixed $n_\mathrm{e}$ covers a wide range of these strong-line ratios and form a narrow sequence parallel to the ridge line of the contours.  The trend in the $N_2$ vs. $O_3$ diagram was reported by \citet{2008MNRAS.385..769B}.

\bsp	
\label{lastpage}
\end{document}